\documentclass[apj]{emulateapj}
\usepackage{subfigure}
\usepackage{graphics}

\def\spose#1{\hbox to 0pt{#1\hss}}
\def\simlt{\mathrel{\spose{\lower 3pt\hbox{$\mathchar"218$}}
    \raise 2.0pt\hbox{$\mathchar"13C$}}}
\def\simgt{\mathrel{\spose{\lower 3pt\hbox{$\mathchar"218$}}
    \raise 2.0pt\hbox{$\mathchar"13E$}}}

\newcommand{\oiii}{\mbox{[\ion{O}{3}]} $\,$}

\newcommand{\oiiiw}{\mbox{[\ion{O}{3}] $\lambda$5007} $\,$}
\newcommand{\oiiiwn}{\mbox{[\ion{O}{3}] $\lambda$5007}}
\newcommand{\nii}{\mbox{[\ion{N}{2}]} $\,$}

\newcommand{\ha}{\mbox{H$\alpha$} $\,$}
\newcommand{\han}{\mbox{H$\alpha$}}

\newcommand{\oiiihb}{\mbox{[\ion{O}{3}] $\lambda$5007}/{\mbox{H$\beta$} $\,$}}
\newcommand{\oiiihbn}{\mbox{[\ion{O}{3}] $\lambda$5007}/{\mbox{H$\beta$}}}

\newcommand{\niiha}{\mbox{[\ion{N}{2}]}/{\mbox{H$\alpha$}} $\,$}
\newcommand{\niihan}{\mbox{[\ion{N}{2}]}/{\mbox{H$\alpha$}}}
\newcommand{\oiiis}{\mbox{[\ion{O}{3}] $\lambda$4363} $\,$}
\newcommand{\oiiidouble}{\mbox{[\ion{O}{3}] $\lambda\lambda$4959, 5007}}
\newcommand{\oidouble}{\mbox{[\ion{O}{1}] $\lambda\lambda$6300, 6363}}
\newcommand{\niidouble}{\mbox{[\ion{N}{2}] $\lambda\lambda$6548, 6583}}
\newcommand{\oiha}{\mbox{[\ion{O}{1}] $\lambda$6300/H$\alpha$} $\,$} 
\newcommand{\heiihb}{\mbox{\ion{He}{2} $\lambda$4686/H$\beta$} $\,$}
\newcommand{\heiihbn}{\mbox{\ion{He}{2} $\lambda$4686/H$\beta$}}
\newcommand{\siisii}{\mbox{[\ion{S}{2}] $\lambda\lambda$6717/6731} $\,$}

\slugcomment{\it ApJ accepted}
\shortauthors{Comerford et al.}
\shorttitle{An Active Galactic Nucleus Caught in the Act of Turning Off and On}

\begin{document}
  
\title{An Active Galactic Nucleus Caught in the Act of Turning Off and On} 

\author{Julia M. Comerford\altaffilmark{1}, R. Scott Barrows\altaffilmark{1}, Francisco M\"{u}ller-S\'{a}nchez\altaffilmark{1}, Rebecca Nevin\altaffilmark{1}, Jenny E. Greene\altaffilmark{2}, David Pooley\altaffilmark{3}, Daniel Stern\altaffilmark{4}, and Fiona A. Harrison\altaffilmark{5}}

\affil{$^1$Department of Astrophysical and Planetary Sciences, University of Colorado, Boulder, CO 80309, USA}
\affil{$^2$Department of Astrophysical Sciences, Princeton University, Princeton, NJ 08544, USA}
\affil{$^3$Department of Physics and Astronomy, Trinity University, San Antonio, TX 78212, USA}
\affil{$^4$Jet Propulsion Laboratory, California Institute of Technology, 4800 Oak Grove Drive, Pasadena, CA 91109, USA}
\affil{$^5$California Institute of Technology, 1200 East California Boulevard, Pasadena, CA 91125, USA}

\begin{abstract}
We present the discovery of an active galactic nucleus (AGN) that is turning off and then on again in the $z=0.06$ galaxy SDSS J1354+1327.  This episodic nuclear activity is the result of discrete accretion events, which could have been triggered by a past interaction with the companion galaxy that is currently located 12.5 kpc away.  We originally targeted SDSS J1354+1327 because its Sloan Digital Sky Survey spectrum has narrow AGN emission lines that exhibit a velocity offset of 69 km s$^{-1}$ relative to systemic.  To determine the nature of the galaxy and its velocity-offset emission lines, we observed SDSS J1354+1327 with {\it Chandra}/ACIS, {\it Hubble Space Telescope}/Wide Field Camera 3, Apache Point Observatory optical longslit spectroscopy, and Keck/OSIRIS integral-field spectroscopy.  We find a $\sim10$ kpc cone of photoionized gas south of the galaxy center and a $\sim1$ kpc semi-spherical front of shocked gas, which is responsible for the velocity offset in the emission lines, north of the galaxy center.  We interpret these two outflows as the result of two separate AGN accretion events; the first AGN outburst created the southern outflow, and then $<10^5$ yrs later the second AGN outburst launched the northern shock front.  SDSS J1354+1327 is the galaxy with the strongest evidence for an AGN that has turned off and then on again, and it fits into the broader context of AGN flickering that includes observations of AGN light echoes.
\end{abstract}

\keywords{ galaxies: active -- galaxies: nuclei }

\section{Introduction}
\label{intro}

When a supermassive black hole (SMBH) accretes a sufficient amount of gas, it shines as an active galactic nucleus (AGN).  However, this is not a continuous process; SMBHs can turn on and off as AGNs depending on their accretion histories.  A local example is the SMBH in the Milky Way, which is currently quiescent, but diffuse bubbles of gamma-ray emission extending $\sim10$ kpc above and below the Galactic plane indicate that the SMBH was active $<10^7$ yrs ago \citep{SU10.1}.  Evidence of this past activity is also seen in X-ray reflection off galactic center molecular clouds (e.g., \citealt{KO96.2, ZH15.1}).

Typical AGN variability timescales have been inferred to be $10^4$ to $10^5$ yrs between turning on and turning off (e.g., \citealt{SC15.1}).  This time variability is linked to accretion of clumpy interstellar medium (ISM) in individual events, as seen in simulations of SMBH accretion and the corresponding energetic AGN outbursts (e.g., \citealt{HO10.1,GA13.1,GA14.2}).  Since observable emission signatures can linger for up to $10^5$ yrs after an AGN has turned off in a galaxy, this leads to observable light echoes from past AGN activity, which are seen in systems such as voorwerpen and changing-look AGNs (e.g., \citealt{TR07.1,LI09.2,FO14.1,KE15.1,LA15.2,RU16.1,SC16.2}).

Here we present a galaxy that has evidence of two discrete AGN outburst events.  SDSS J135429.05+132757.2 (hereafter SDSS J1354+1327, located at $z=0.06$) was originally noted for its velocity-offset narrow emission lines.  By analogy to double-peaked narrow emission lines used to identify dual AGN candidates (e.g., \citealt{LI10.1,CO11.2,FU12.1,BA12.1,BA13.1,MC15.1,MU15.1}), single-peaked narrow AGN emission lines that exhibit a velocity offset relative to systemic can be used to select offset AGN candidates \citep{CO09.1,CO13.1}.  Follow-up observations of some of these galaxies that exhibit velocity-offset AGN emission lines in their Sloan Digital Sky Survey (SDSS) spectra have revealed that the velocity-offset emission lines can be produced by offset AGNs or by shocked gas in outflows or inflows \citep{AL15.1,BA16.1,MU16.1}.

\begin{figure*}
\begin{center}
\includegraphics[width=3.2in]{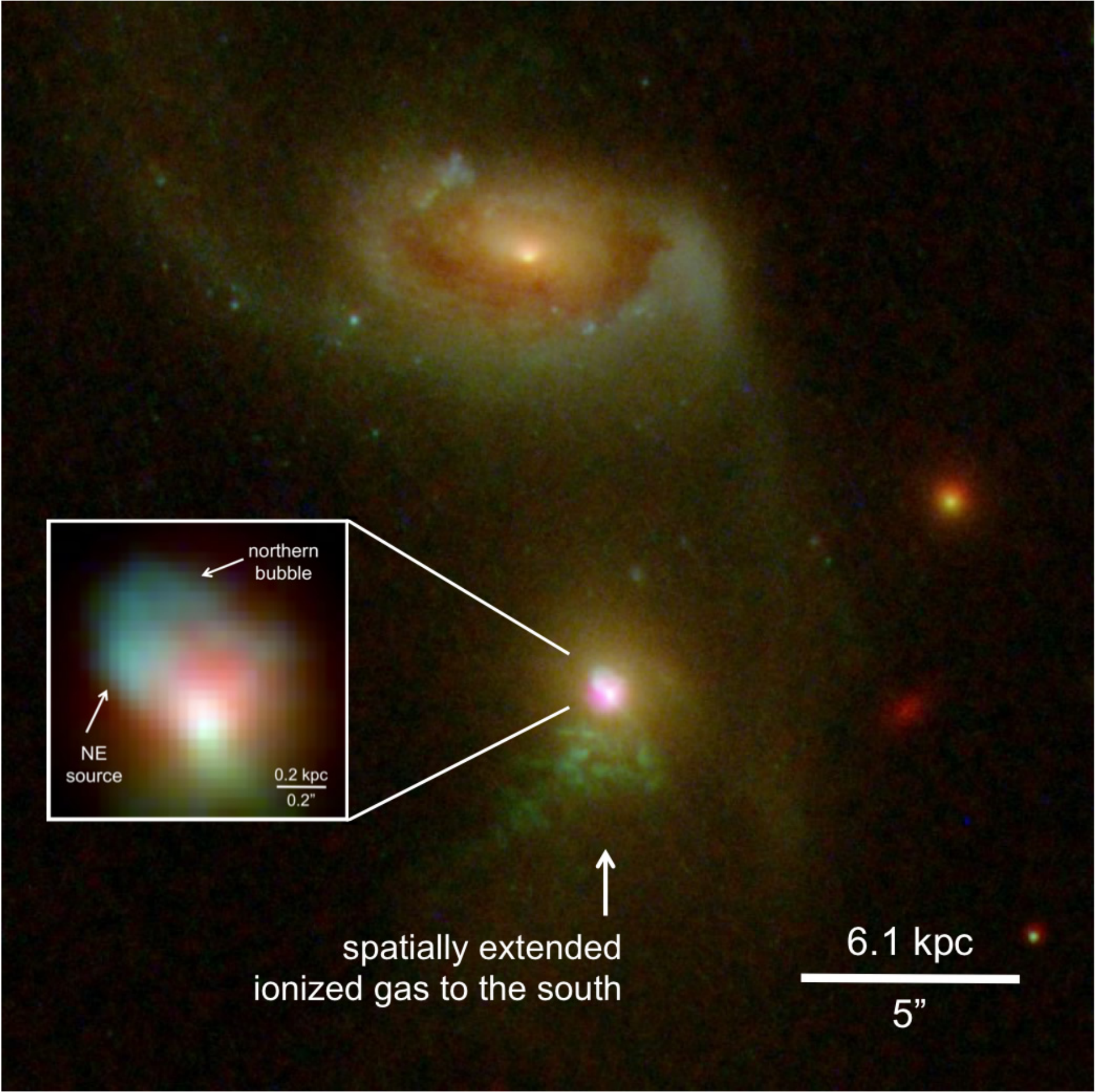}
\includegraphics[width=3.2in]{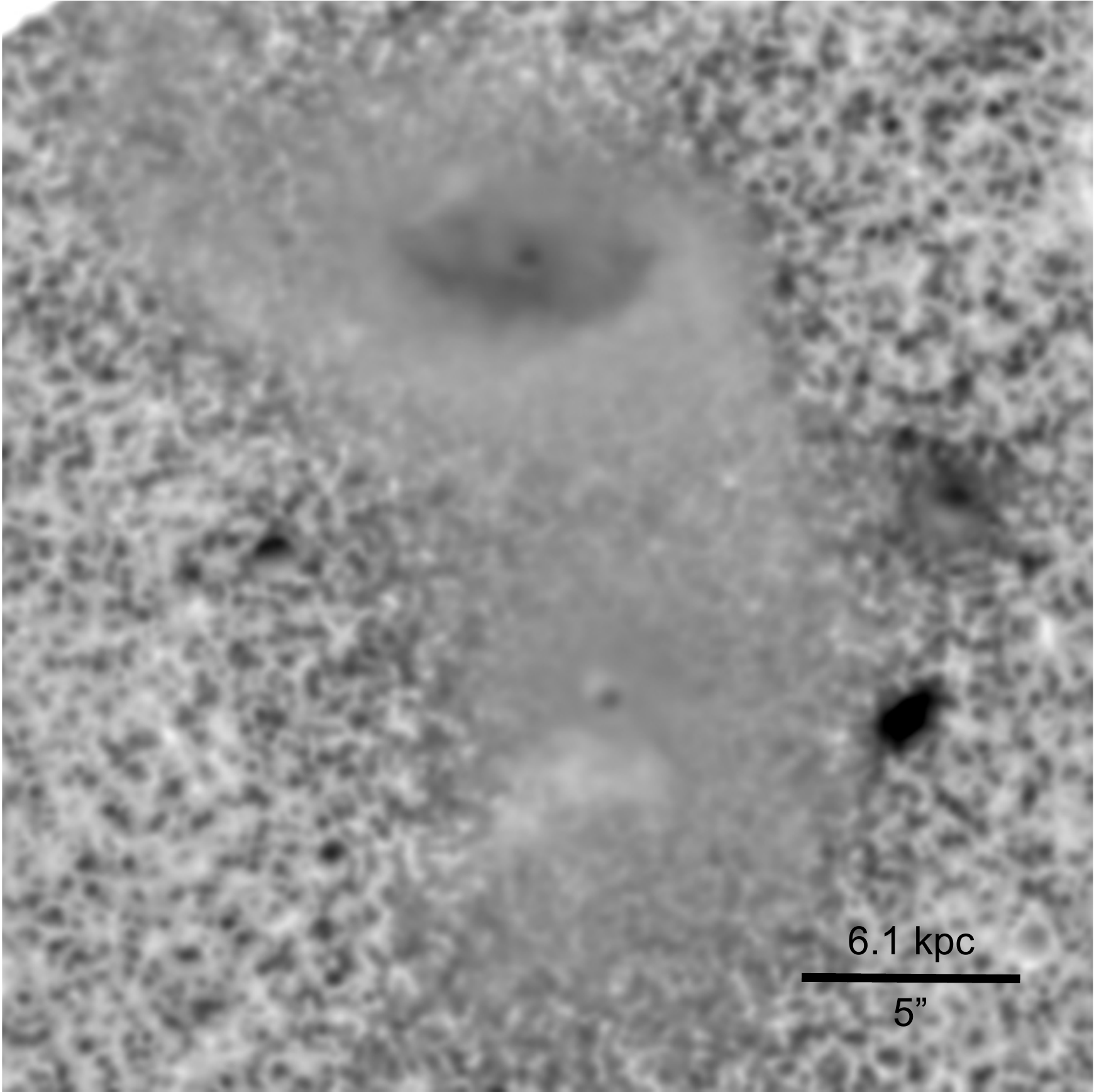}
\end{center}
\caption{$25^{\prime\prime} \times 25^{\prime\prime}$ images of SDSS J1354+1327 and its companion galaxy SDSS J1354+1328, which is located 12.5 kpc to the northeast.  Left: a four-color image, where the observations shown are {\it HST} F160W (red), F606W (green), F438W (blue), and {\it Chandra} restframe $0.5-10$ keV (purple, one-twelfth size pixels smoothed with a 16 pixel radius Gaussian kernel).  The inset shows a $1\farcs2 \times 1\farcs2$ zoom around the northern region of the galaxy, with only the {\it HST} data shown to more clearly illustrate the northern bubble of ionized gas. Right: a $V-H$ color map to illustrate the distribution of dust in the system.  The color map is plotted on a logarithmic gray scale, where dark corresponds to redder colors and light corresponds to bluer colors.  In all images, north is up and east is to the left.}
\label{fig:1354companion}
\end{figure*}

To determine the nature of SDSS J1354+1327, we have observed the galaxy with the {\it Chandra X-ray Observatory}, {\it Hubble Space Telescope} ({\it HST}) imaging, Apache Point Observatory Dual Imaging Spectrograph (APO/DIS) optical longslit spectroscopy, and adaptive optics-assisted Keck OH-Suppressing Infra-Red Imaging Spectrograph (Keck/OSIRIS) integral-field spectroscopy (IFS).  We conclude that SDSS J1354+1327 is the best observational example of an AGN turning off and then on, with a relic outflow to the south and a new AGN outburst north of the galaxy center (Figure~\ref{fig:1354companion}, left).  We find that the velocity-offset emission lines are produced by shocks emanating from the new AGN outburst.  

The paper is organized as follows: in Section~2 we present our observations, including space-based and ground-based observations of SDSS J1354+1327 and analyses of these data.  In Section~3 we combine the observations to interpret the nature of the emission sources in SDSS J1354+1327 and the nature of its companion galaxy.  In Section~4 we determine the nature of SDSS J1354+1327 itself.  In Section~5 we present our conclusions.  We assume a Hubble constant $H_0 =70$ km s$^{-1}$ Mpc$^{-1}$, $\Omega_m=0.3$, and $\Omega_\Lambda=0.7$ throughout, and all distances are given in physical (not comoving) units.

\section{Observations and Analysis}

\subsection{Optical SDSS Observations}
\label{sdss}

The spectrum of SDSS J1354+1327 is classified as a Type 2 AGN and has no evidence of broad emission lines \citep{OH15.1}.  We obtain the fluxes of the emission lines in the SDSS spectrum of the galaxy using the OSSY catalog \citep{OH11.1}, which simultaneously fits an entire spectrum using stellar templates for the stellar kinematics and Gaussian templates for the emission components.  The host galaxy redshift, based on the template fit to the stellar absorption features, is $z=0.063316 \pm 0.000070$, which we use to convert the \oiiiw flux to a luminosity $L_{\oiiiwn}=4.56 \times 10^{41}$ erg s$^{-1}$.  The \oiiiw emission line is redshifted by $69.1 \pm 15.0$ km s$^{-1}$ relative to the systemic velocity of the galaxy's stars \citep{CO14.1}.  We find that the line flux ratios are $\log (\oiiihbn)=1.03 \pm 0.05$ and $\log (\niihan)=-0.50 \pm 0.05$, which place SDSS J1354+1327 in the Seyfert region of the Baldwin-Phillips-Terlevich (BPT) diagram \citep{BA81.1,KE06.1}.  
 
The companion galaxy SDSS J1354+1328 has a redshift of $z=0.063586 \pm 0.000048$ and an \oiiiw luminosity of $L_{\oiiiwn}=2.4 \times 10^{40}$ erg s$^{-1}$.  The line flux ratios are $\log (\oiiihbn)=0.44 \pm 0.09$ and $\log (\niihan)=-0.26 \pm 0.08$, which are in the Seyfert region of the BPT diagram.  

\begin{figure*}
\begin{center}
\includegraphics[width=6.7in]{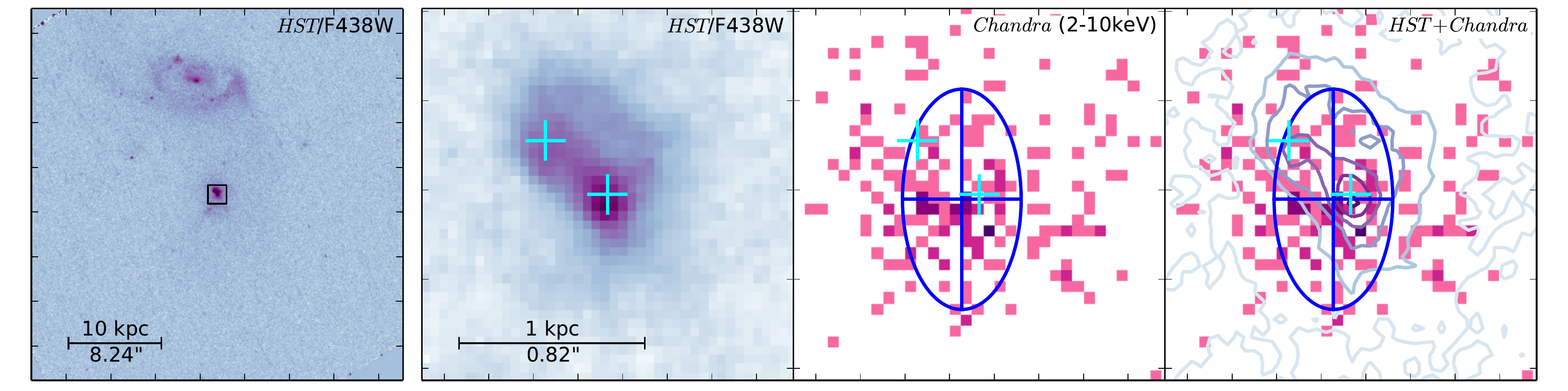}
\includegraphics[width=6.7in]{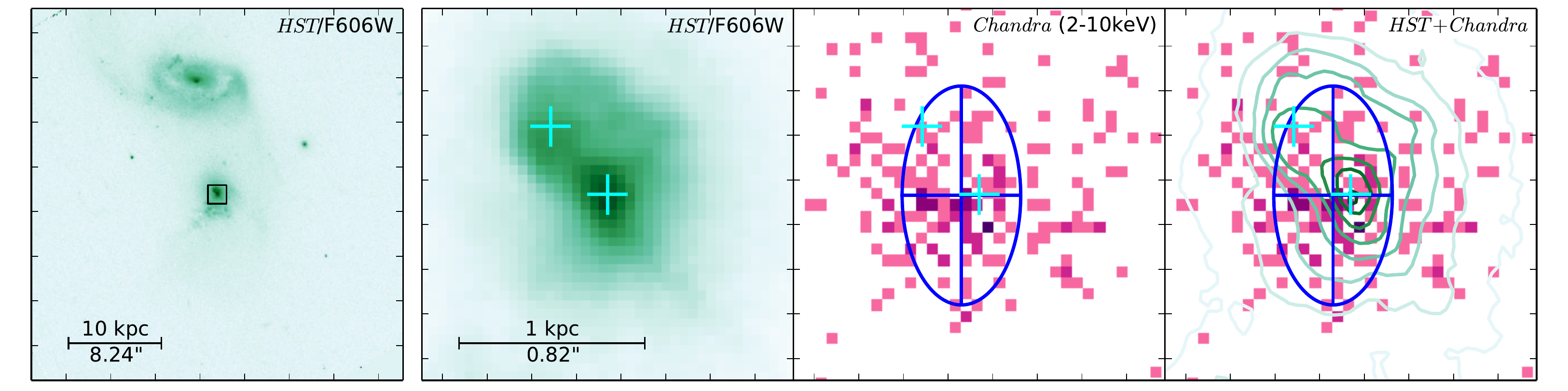}
\includegraphics[width=6.7in]{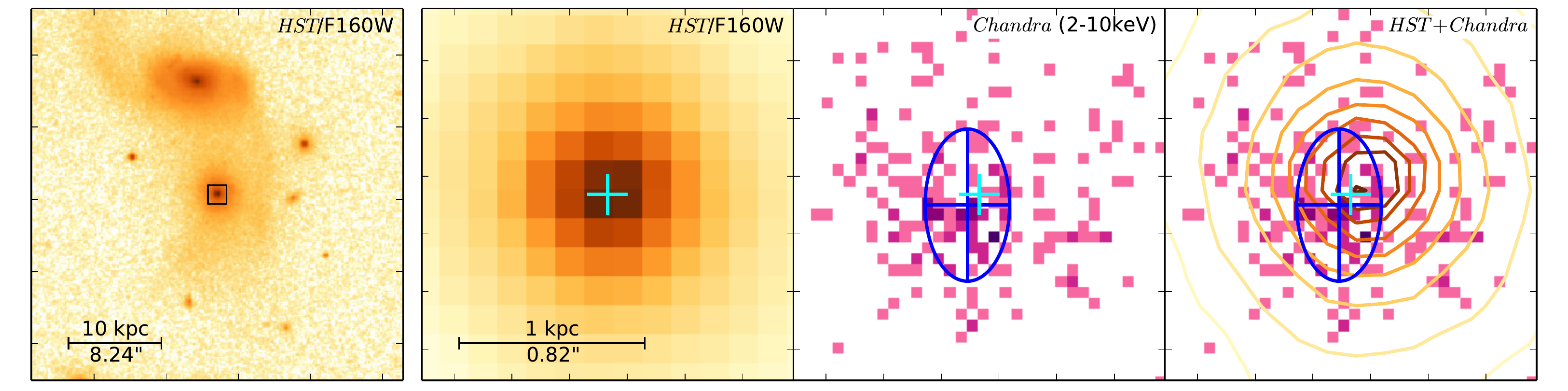}
\end{center}
\caption{Top to bottom: F438W, F606W, and F160W observations of SDSS J1354+1327.  {\bf Left}: {\it HST} image, with a box illustrating the region of SDSS J1354+1327 shown in the figures to the right.  {\bf Middle left}: $2^{\prime\prime} \times 2^{\prime\prime}$ {\it HST} image, with cyan crosses marking the central emission source and the NE emission source (or in the case of F160W, marking only the stellar centroid).  {\bf Middle right}: $2^{\prime\prime} \times 2^{\prime\prime}$ image of restframe $2-10$ keV {\it Chandra} observations, where the blue cross marks the position of the X-ray source and the size of the cross illustrates the $3\sigma$ errors on the X-ray source position, given the relative astrometric uncertainties between the {\it Chandra} image and the {\it HST} band shown.  {\bf Right}: $2^{\prime\prime} \times 2^{\prime\prime}$ image of restframe $2-10$ keV {\it Chandra} observations, with {\it HST} contours overlaid.  The cyan and blue crosses are as shown in the middle left and middle right figures, respectively.  In all panels, north is up and east is to the left.}
\label{fig:positions}
\end{figure*}

\subsection{Chandra/ACIS X-ray Observations}
\label{chandra}

SDSS J1354+1327 was observed with {\it Chandra}/ACIS for 9443 seconds on UT 2014 June 25, as part of the program GO4-15113X (PI: Comerford).  The data were taken with the telescope aimpoint on the ACIS S3 chip in ``timed exposure'' mode and telemetered to the ground in ``faint'' mode.  We reduced the data with the latest {\it Chandra} software (CIAO\,4.6.1) and the most recent set of calibration files (CALDB\,4.6.2).

We used \texttt{dmcopy} to make sky images of the field in the soft ($S$, restframe $0.5-2$ keV), hard ($H$, restframe $2-10$ keV) and total ($T$, restframe $0.5-10$ keV) energy ranges with pixels binned to 1/10th the native pixel size.  Then, we used \texttt{Sherpa} to model the X-ray source as a two-dimensional Lorenztian function in \texttt{beta2d}.  Our model also included a background component of fixed count rate, that we determined from an adjacent circular region of $30^{\prime\prime}$ radius.  We confirmed that the background region does not contain any sources detected by \texttt{wavdetect} with a threshold of \texttt{sigthresh}$=10^{-8}$.  We set the initial position of the \texttt{beta2d} component to the location of the SDSS galaxy coordinates. After using \texttt{psfSize} to estimate that the radius of the point spread function (PSF) is $1\farcs07$, we ran the fit in \texttt{Sherpa} and allowed the model to fit a region around the galaxy coordinates of 3 times the PSF size at that location on the chip. We determined the best-fit model parameters by minimizing the Cash statistic using  \texttt{Sherpa}'s implementation of the 'Simplex' minimization algorithm \citep{LA98.1}.  To test for additional sources, we also attempted fitting a two-component \texttt{beta2d} model.  However, the amplitude of any second component is detected at $<1\sigma$ significance, and therefore we adopt the single \texttt{beta2d} model.  

To determine the relative positions of the {\it Chandra} sources and the {\it HST} sources, we registered the {\it Chandra} and {\it HST}/F160W images and estimated the relative astrometric uncertainty between the two images.  Due to the small number of {\it Chandra} sources and the relatively small {\it HST}/F160W field of view, we could not directly register the two images.  Instead, we registered both images with external images: SDSS ($u$, $g$, $r$, $i$, and $z$) and the 2MASS point source catalog \citep{CU03.1}.   This provided us with six independent estimates of the transformations and relative astrometric uncertainty between the {\it Chandra} and {\it HST} images.  Finally, to minimize the relative astrometric uncertainty, we computed an error-weighted average of the six independent transformations.  In the {\it Chandra} physical ($X$,$Y$) coordinate system, the error-weighted averages of these six transformations yield the final astrometric errors $\overline{\Delta X}=0\farcs0583$ and $\overline{\Delta Y}=0\farcs1104$.

The best-fit positions of the soft, hard, and total X-ray sources, with astrometric shifts applied to align the positions in the {\it HST} coordinate frame, are (13:54:29.052, +13:27:56.70), (13:54:29.050, +13:27:56.81), and (13:54:29.047, +13:27:56.84), respectively.  The errors on the positions are ($0\farcs19$, $0\farcs50$), ($0\farcs07$, $0\farcs27$), and ($0\farcs06$, $0\farcs20$), respectively.  The X-ray positions are shown in Figure~\ref{fig:positions}.
 
Next, we measured the numbers of soft, hard, and total counts attributed to the X-ray source using the Bayesian Estimation of Hardness Ratios (\texttt{BEHR}) code described in \citep{PA06.1}.  \texttt{BEHR} takes as input the observed soft and hard counts from both the source region and a background region (which we measured using \texttt{calc$\_$data$\_$sum}).  The code then uses a Bayesian approach to estimate the expected values and uncertainties of the soft counts, hard counts, total counts, and hardness ratio.  Using this approach, we found $S=13.7^{+3.1}_{-4.2}$ counts, $H=238^{+13.9}_{-17.0}$ counts, and $T=252^{+14.8}_{-17.0}$ counts. 

We then used \texttt{Sherpa} to model the unbinned energy spectrum of the extracted region over the observed energy range of 2-8 keV. For our purposes, we are interested in the observed and intrinsic fluxes integrated over restframe soft, hard, and total energy ranges. Therefore, we fit the spectrum with a redshifted power law, $F=E^{-\Gamma}$, (intended to represent the intrinsic AGN X-ray emission at the SDSS spectroscopic redshift, $z_{\rm{SDSS}}$).  We also included two multiplicative model components of absorbing neutral Hydrogen column densities.  The first absorbing component is fixed to the Galactic value, $n_{\rm{H,gal}}$, which we determined using an all-sky interpolation of the neutral Hydrogen in the Galaxy \citep{DI90.1}.  The second absorbing component is allowed to vary and assumed to be intrinsic to the source, $n_{\rm{H,exgal}}$, at the redshift $z_{\rm{SDSS}}$.  All fluxes are $k$-corrected, with the observed values calculated from the model sum (including the absorbing components), while the intrinsic values are calculated from the unabsorbed power law component.  Since we have used Cash statistics, we did not subtract the background before modeling, but we have confirmed that the counts over the 2-8 keV range are dominated by the source region, with a negligible contribution from the background region.

\begin{figure}[!h]
\begin{center}
\includegraphics[width=3.5in]{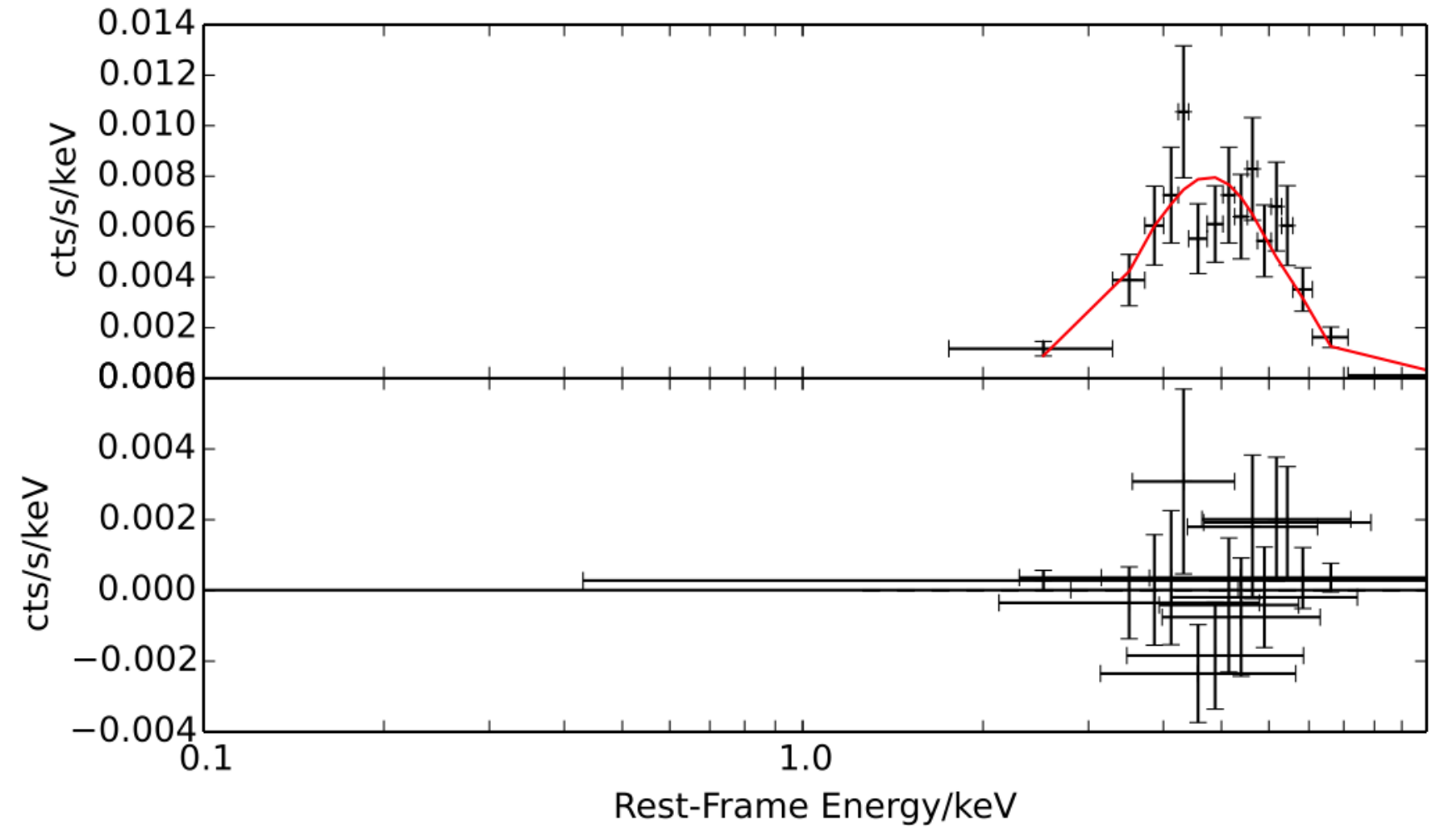}
\end{center}
\caption{X-ray spectral fit to SDSS J1354+1327, where the top panel shows the fit in red and the bottom panel shows the residuals.}
\label{fig:chandraspectrum}
\end{figure}

For our first fit to the spectrum, we allowed $\Gamma$ and $n_{\rm{H,exgal}}$ to vary freely.  We found a best-fit value of $\Gamma=-1.81$ (and a corresponding upper limit on the column density of $3.4 \times 10^{21}$ cm$^{-2}$), which is not within the typical range of observed power-law indices for AGNs, i.e. $1\le \Gamma \le 3$ \citep{IS10.1,PI05.1,NA94.2,RE00.1}.  Consequently, we fixed $\Gamma$ at a value of 1.70 and ran the fit again.  We obtained the best-fit model parameters for each combination of parameters by minimizing the Cash statistic using \texttt{Sherpa}'s implementation of the Levenberg-Marquardt optimization method \citep{BE69.1}.  The best-fit column density is $1.96^{+0.29}_{-0.23} \times 10^{23}$ cm$^{-2}$, where the errors do not account for the uncertainty in $\Gamma$, and the reduced Cash statistic is 0.94. Figure~\ref{fig:chandraspectrum} shows the best fit to the spectrum. 

The restframe hard absorbed (unabsorbed) X-ray fluxes are $1.1^{+0.2}_{-0.1} \times 10^{-12}$ erg s$^{-1}$ cm$^{-2}$ ($2.8^{+0.3}_{-0.3} \times 10^{-12}$ erg s$^{-1}$ cm$^{-2}$) for 2-10 keV.  We then used the redshift $z_{\rm{SDSS}}$ to convert the X-ray fluxes to X-ray luminosities. The restframe hard absorbed (unabsorbed) luminosities are $L_{X, 2-10 \mathrm{keV}}=1.1^{+0.2}_{-0.1} \times 10^{43}$ erg s$^{-1} (2.7^{+0.3}_{-0.3} \times 10^{43}$ erg s$^{-1})$.

Finally, we searched for an X-ray source at the center of the companion galaxy SDSS J1354+1328, but found none (detection of $0.001^{+0.002}_{-0.001}$ counts in the restframe $0.5-10$ keV energy range).

\begin{figure*}[!t]
\begin{center}
\includegraphics[width=6.5in]{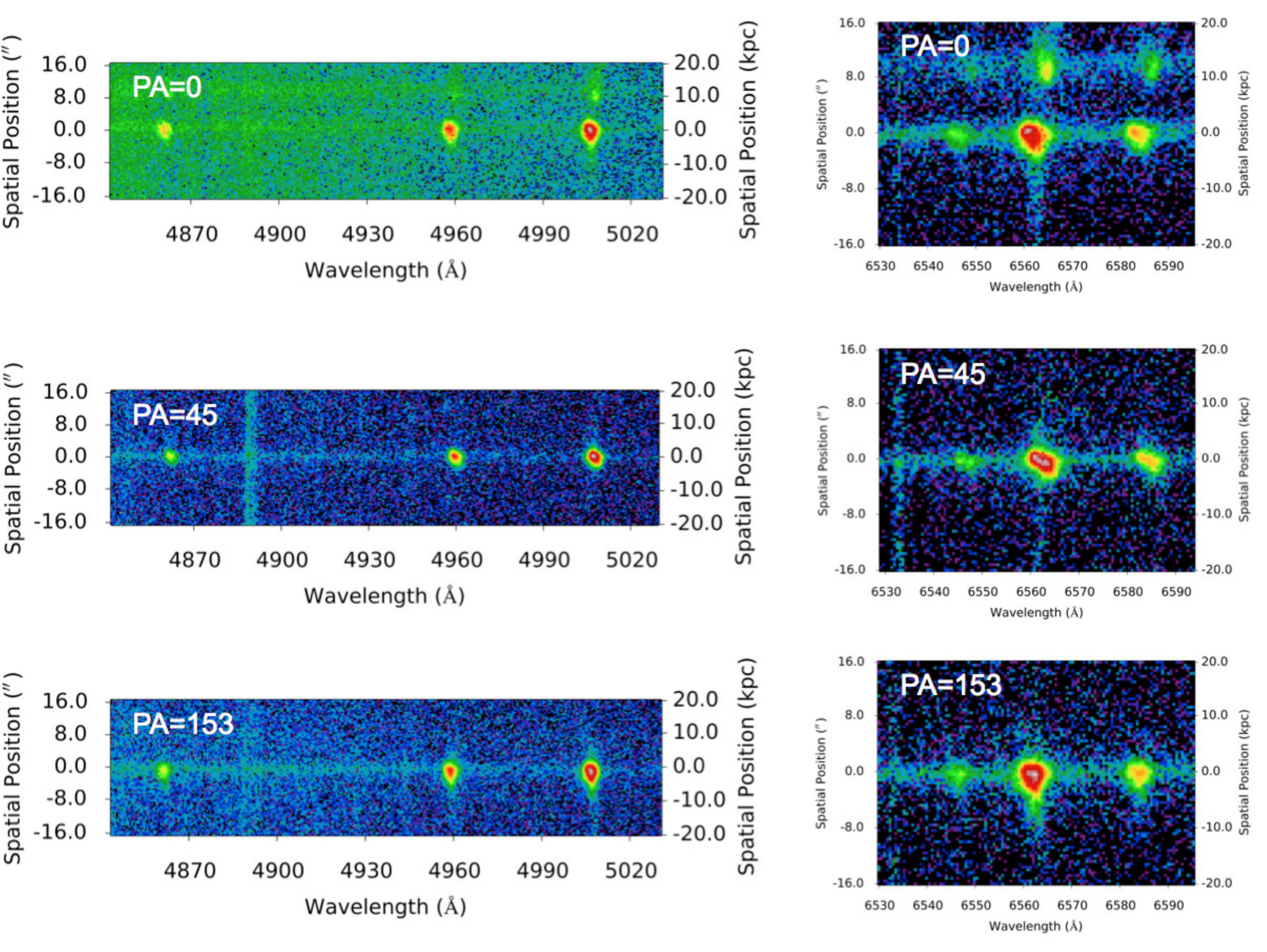}
\end{center}
\caption{Segments of the two-dimensional APO/DIS longslit spectra taken at three different position angles: 0 (top; the companion galaxy is visible above the target galaxy), 45 (middle), and 153 (bottom) degrees East of North.  The blue channel spectra are shown on the left, and the red channel spectra are shown on the right.  The wavelengths shown are restframe wavelengths based on the galaxy's stellar absorption features.  The longslit spectra are centered spatially on the galaxy continuum, with positive spatial positions to the north and negative spatial positions to the south. }
\label{fig:longslit}
\end{figure*}

\subsection{HST/WFC3 F160W, F606W, and F438W  Observations}
\label{hst}

SDSS J1354+1327 was also observed with {\it HST}/WFC3 on UT 2014 April 28 (GO 13513, PI:Comerford), and the observations covered three bands: UVIS/F438W ($B$ band, 954 seconds), UVIS/F606W ($V$ band, 900 seconds), and IR/F160W ($H$ band, 147 seconds).  The {\it HST} field of view included SDSS J1354+1328, which is the companion galaxy to the northeast of SDSS J1354+1327.  We used GALFIT V3.0 \citep{PE10.1} to model the light profiles of these two galaxies.  GALFIT is capable of decomposing images of galaxies into multiple components, including a S\'ersic profile (e.g., a galactic stellar bulge), exponential disk (e.g., a galactic disk), and an image PSF for bright unresolved sources (e.g., AGNs).  

We fit a S\'ersic profile (plus a fixed, uniform sky component) to locate the position of each central stellar bulge, since S\'ersic profiles have been empirically shown to be good approximations of the light profiles of stellar bulges \citep{GR05.2}. To avoid modeling complex and irregular gas kinematics, we only ran GALFIT models on the F160W image because it does not contain significant line emission from ionized gas and it is sensitive to the central stellar bulge, which is the primary component of interest.  As shown in \citet{CO15.1}, an additional unresolved point source is not necessary in this model since the AGN contributes negligibly to the F160W flux.  The fit was run on a square region of projected physical size 40 kpc on each side, in order to explore within 20 kpc of the AGN position.  

We find that the position of SDSS J1354+1327's central stellar bulge is (13:54:29.045, +13:27:56.88), with errors of ($0\farcs$06, $0\farcs11$). The separation between SDSS J1354+1327 and the companion galaxy is $12.5$ kpc ($10\farcs2$).  Using the results of the best-fit GALFIT model to each galaxy, we also find the $H$-band luminosity for SDSS J1354+1327 ($L_H=4.0 \times 10^{10}$ $L_\odot$) and for the companion galaxy ($L_H=4.1 \times 10^{10}$ $L_\odot$).

To explore the distribution of dust in the galaxy, we made a $V-H$ color map \citep{MA03.6}.  We convolved the $V$ and $H$ images with a Gaussian kernel to put them at the same angular resolution, and then computed the magnitude difference $V-H$.  The resultant color map is shown in Figure~\ref{fig:1354companion} (right). We find that the center of SDSS J1354+1327 is the most obscured part of the galaxy ($V-H=2$).

Next, we turn our attention to the F438W and F606W observations, which cover a wavelength range that includes $H\delta$, $H\gamma$, and \oiiis (F438W) and H$\beta$, \oiiidouble, \oidouble, \niidouble, and H$\alpha$ (F606W).  The F438W and F606W observations are dominated by ionized gas emission and not stellar emission, since the optical longslit observations show this to be the case for these wavelength ranges and since the morphologies of the F438W and F606W emission are different from the morphology of the stellar continuum (traced in the F160W observations).  We identify two main structures in the F438W and F606W data:

{\it Spatially extended ionized gas to the south.}  South of the galaxy center, there is a cone of clumpy, ionized gas (Figure~\ref{fig:1354companion}, left).  We map the gas detected in F606W at $>10 \sigma$ above the background, and we find that it is distributed in a cone-like structure with an opening angle of $67^\circ$.  The gas detected at $>3 \sigma$ then extends $\sim 10$ kpc radially from the galaxy center.  The range of colors in the cone is $0.2 < V-H < 1.2$, and the patchiness of the color indicates that in some regions the far side of the cone may be obscured. We explore this in more detail in Section~\ref{outflow}.

{\it Northern bubble.}  North of the galaxy center, there is a loop of emission that extends to a maximum distance of $0\farcs6$ (0.7 kpc) from the galaxy center.  The bubble's color is $V-H=0.7$.  By fitting a S\'ersic component, we find that the brightest component of this bubble is located northeast of the galaxy center.  We call this feature the NE source, and in F606W its apparent magnitude is 21.4 (compared to an apparent magnitude of 18.7 for the galaxy).  The positions of the NE source in F438W (13:54:29.064, +13:27:57.10) and F606W (13:54:29.062, +13:27:57.10) agree to within $0\farcs04$, so we take the NE source position to be the average of the positions measured in F438W and F606W.  The NE source is located $0\farcs33$ (0.40 kpc) from the galaxy center. 

\begin{figure}[!t]
\begin{center}
\includegraphics[width=3.5in]{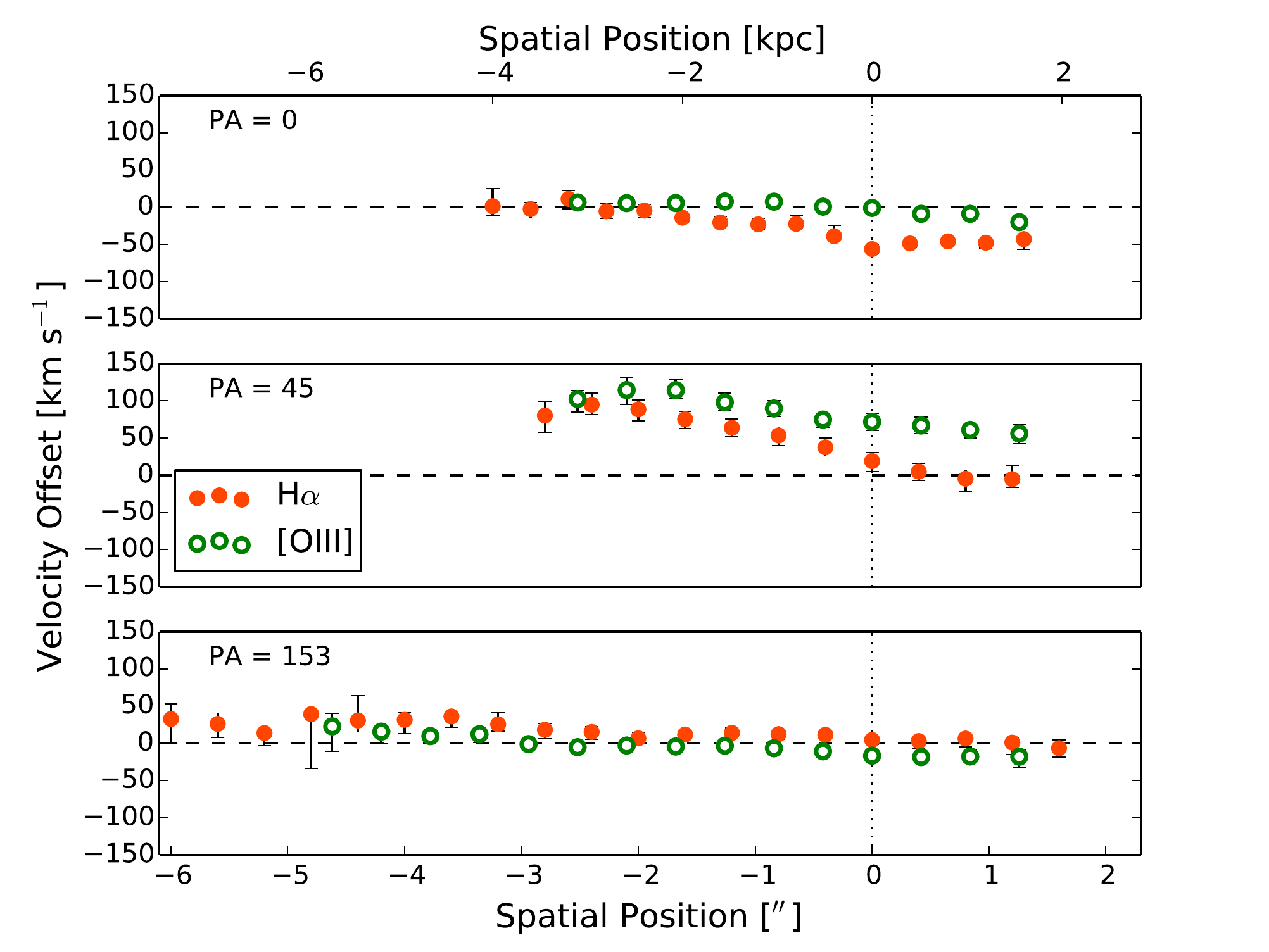}
\end{center}
\caption{Line-of-sight velocity offsets of emission lines along the three different position angles used for the APO/DIS longslit observations.  H$\alpha$ is illustrated with filled red circles, while \oiiiw is illustrated with open green circles, and along each slit the emission detected with signal-to-noise ratios greater than 10 is shown.  A velocity offset of zero (dashed horizontal line) is the systemic velocity of the galaxy's stars.  The spatial position zero marks the centroid of the galaxy continuum (vertical dotted line), with positive spatial positions to the north and negative spatial positions to the south.}
\label{fig:vel}
\end{figure}

\begin{figure*}[!t]
\begin{center}
\includegraphics[width=2.3in]{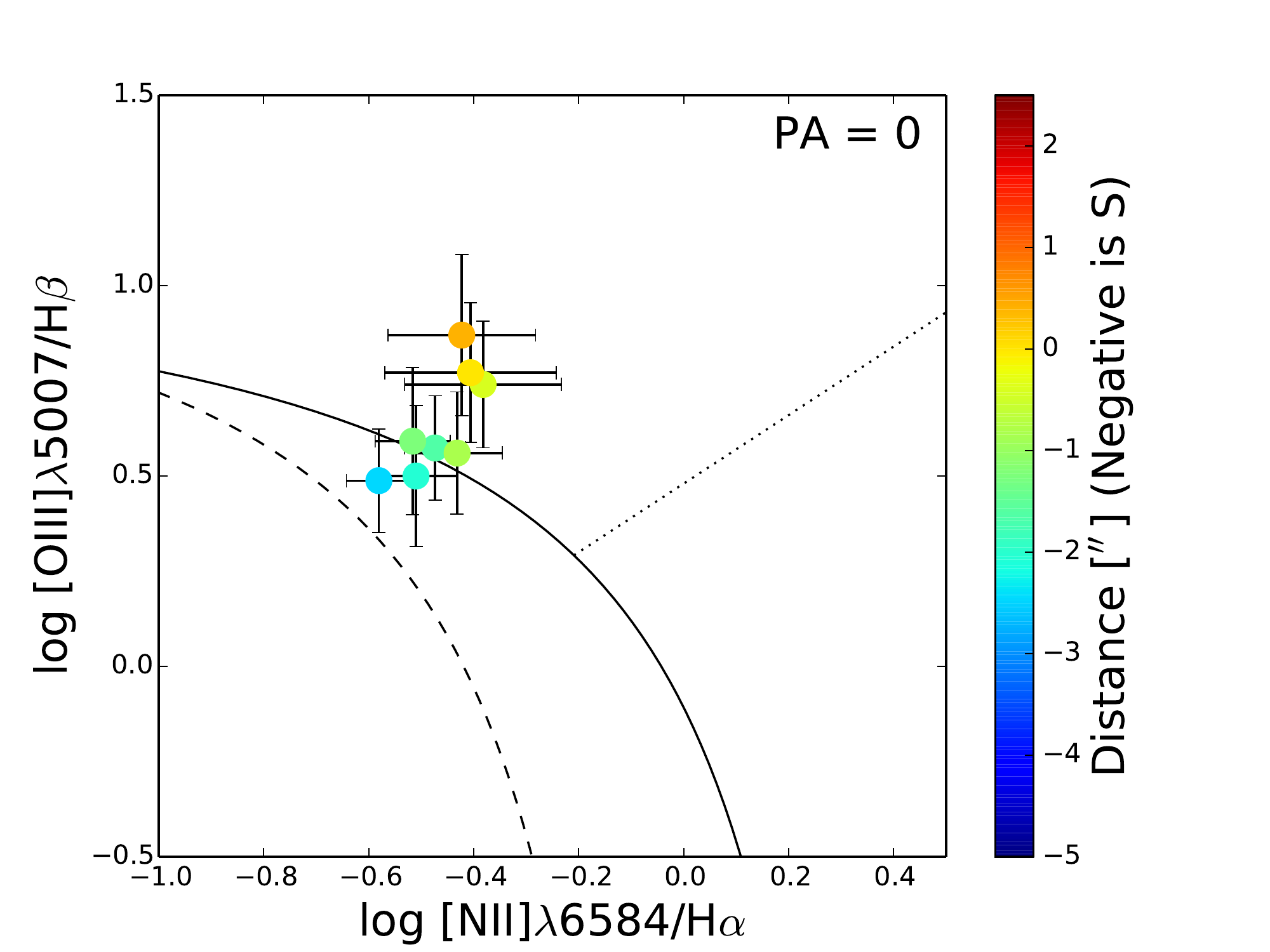}
\hspace{-0.3in}
\includegraphics[width=2.3in]{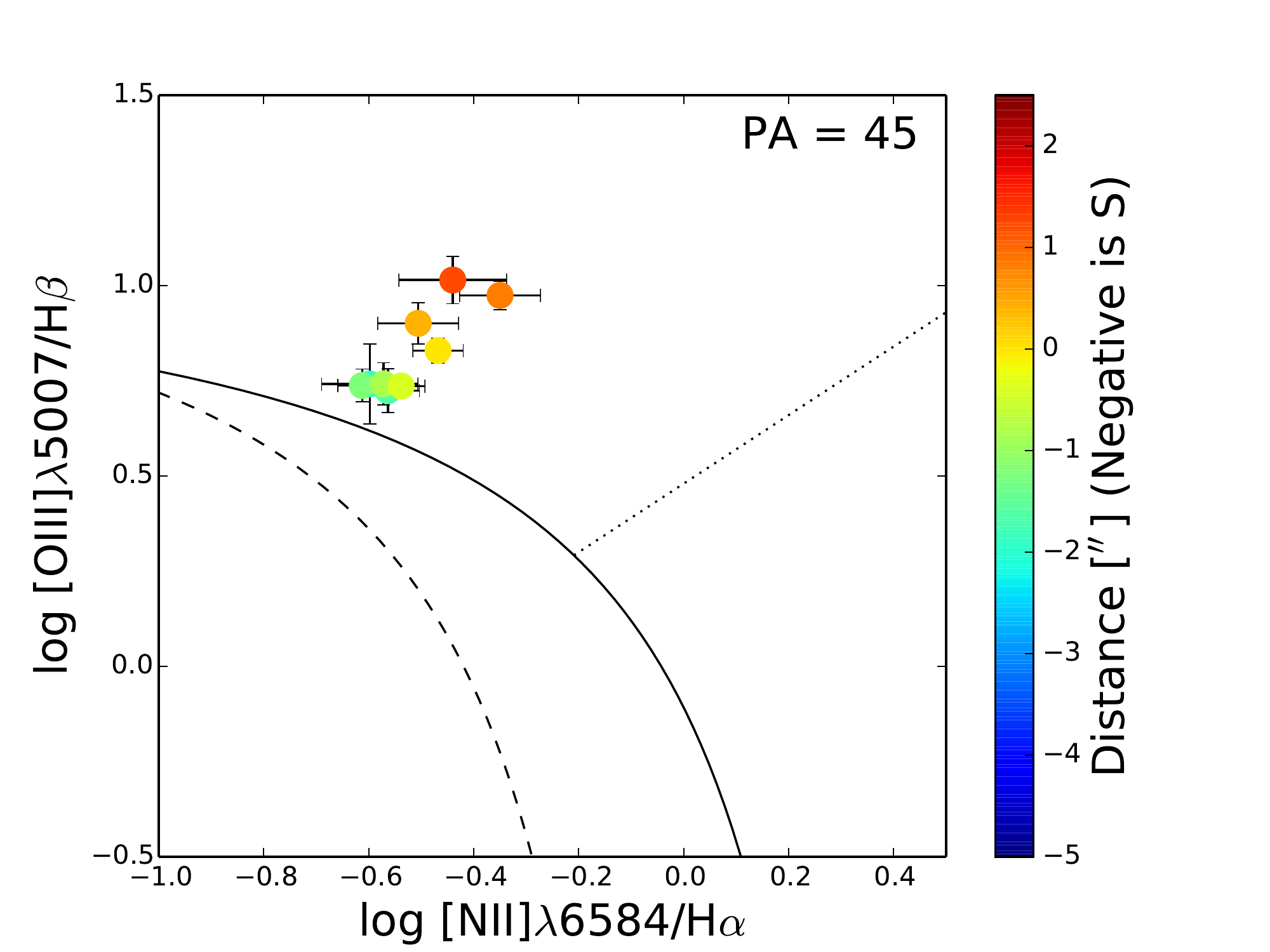}
\hspace{-0.3in}
\includegraphics[width=2.3in]{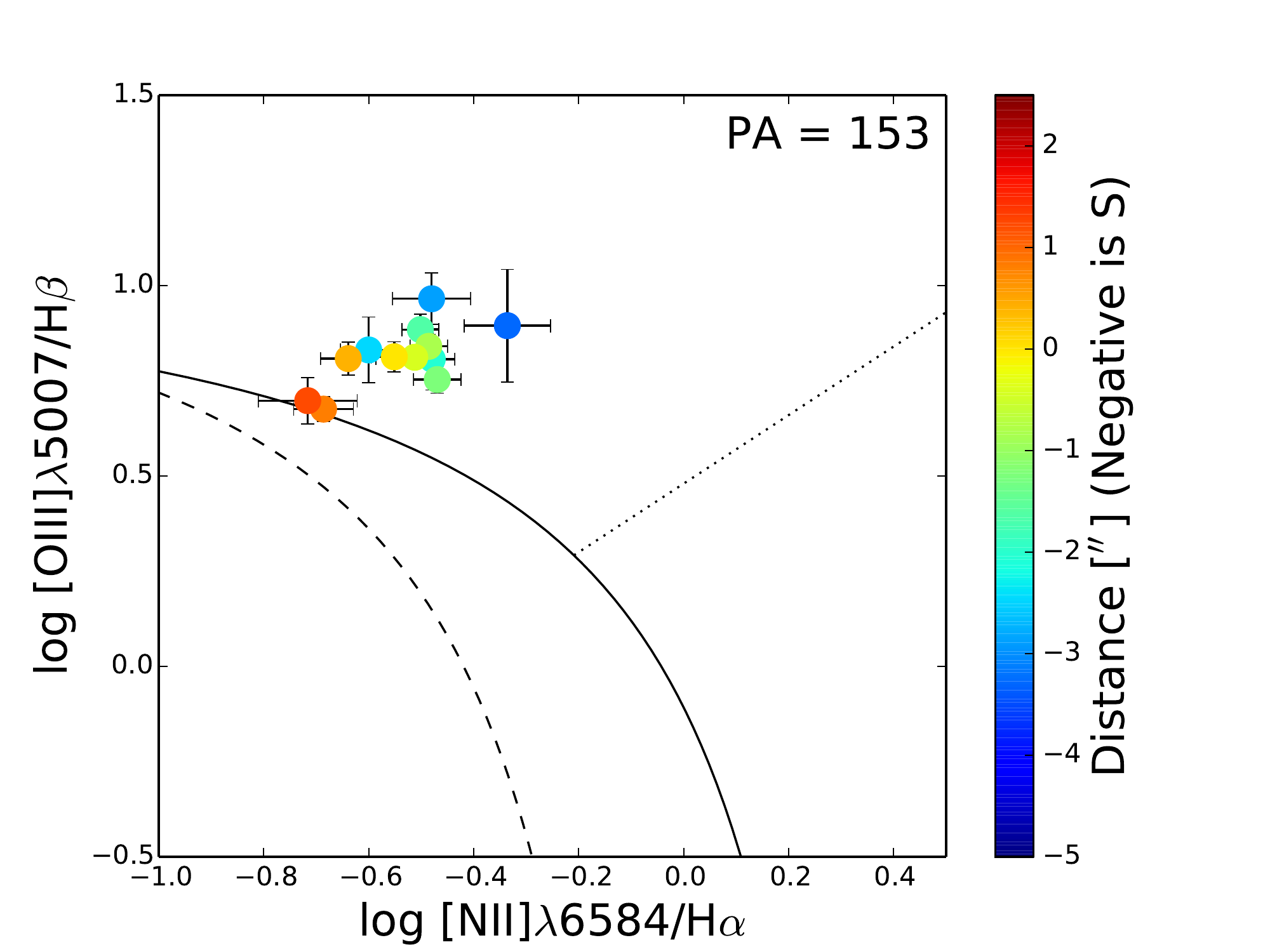}
\end{center}
\caption{BPT diagrams for the three optical longslit position angles: 0 (left), 45 (middle), and 153 (right) degrees East of North.  The solid line shows the theoretical maximum for starbursts \citep{KE01.2}, the dashed line shows the division between purely star-forming and composite (contribution from AGN and star formation) galaxies \citep{KA03.1}, and the dotted line shows the division between Seyferts and LINERs \citep{KE06.1}.  Colored data points and $1\sigma$ error bars are plotted for each position along the slit, where zero is defined as the position of the host galaxy continuum on the slit.}
\label{fig:bpt}
\end{figure*}

\subsection{APO/DIS Optical Longslit Observations}
\label{apo}

We observed SDSS J1354+1327 with APO/DIS on UT 2015 December 18 and UT 2016 June 30. We used a 1200 lines mm$^{-1}$ grating, centered so that the wavelength range covered H$\beta$ and \oiii on the blue channel and \nii and H$\alpha$ on the red channel.  We used a $1\farcs5$ slit width, and the DIS pixel scales are $0\farcs42$ on the blue channel and $0\farcs40$ on the red channel.  Following the approach used to study active galaxies with double-peaked narrow emission lines \citep{CO12.1,NE16.1}, we observed SDSS J1354+1327 at multiple different slit position angles to understand the kinematics of the features of the galaxy that are seen in the {\it HST} images. We observed the galaxy at three different slit position angles (given in degrees East of North): $0^\circ$ to cover the emission seen in the north-south direction and the companion galaxy to the north, $153^\circ$ to cover the bright clumps of emission seen to the south of the galaxy center, and $45^\circ$ to cover the NE emission source.   We observed for 1200s at each position angle, and the data were reduced using standard IRAF procedures.

The longslit spectra are shown in Figure~\ref{fig:longslit}.  We fit Gaussians to the \oiiiw and \ha flux at each position along the slit where the emission line is detected with a signal-to-noise ratio $>10$, and we extracted the line-of-sight velocities.  These are shown in Figure~\ref{fig:vel}.  

Next we identify where the peak of \oiiiw flux is located.  The longslit orientation of position angle 45 has the strongest \oiiiw flux  (integrated flux of $2.6 \times 10^{-13}$ erg s$^{-1}$ cm$^{-2}$, which is twice the integrated flux at position angle 0 and 2.4 times the integrated flux at position angle 153), and we fit a Gaussian to the spatial distribution of \oiiiw flux along this slit position angle.  We find that the peak flux of \oiiiw occurs $0\farcs41 \pm 0\farcs08$ to the north of the galaxy along position angle 45, which is spatially coincident with the NE source seen in the {\it HST} images. The velocity of the \oiiiw peak flux, relative to systemic, is $66.8 \pm 11.5$ km s$^{-1}$.  This explains the velocity offset of $69.1 \pm 15.0$ km s$^{-1}$ seen in the SDSS spectrum of SDSS J1354+1327.

We then use emission line flux ratios in the BPT diagram \citep{BA81.1,VE87.1,KE06.1} to identify the sources of the ionized emission in SDSS J1354+1327.  We measure these line flux ratios at each position along the slit where the signal-to-noise ratio is $>3$ for each of the four emission lines, and the results are shown in Figure~\ref{fig:bpt}.  At all observed position angles the ionized gas has Seyfert-like or composite line ratios. 

\begin{figure}[!h]
\begin{center}
\includegraphics[width=3.5in]{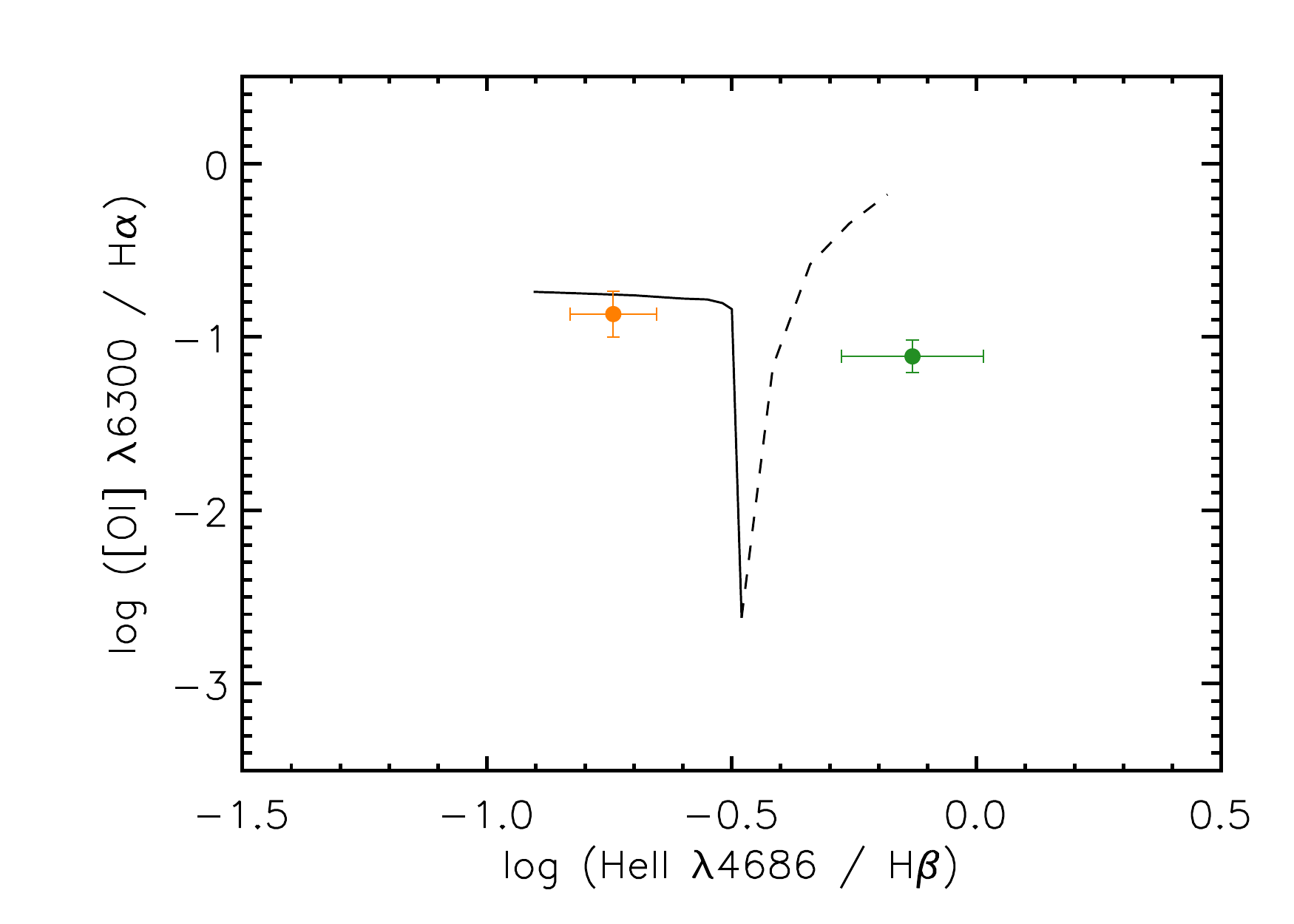}
\end{center}
\caption{\oiha vs. \heiihbn.  The orange point illustrates the line flux ratios of the emission located 3 kpc south of the galaxy center along position angle $153^\circ$, which is part of the spatially extended ionized gas to the south.  The green point illustrates the line flux ratios of the emission located 0.5 kpc north of the galaxy center along position angle $45^\circ$, at the location of the NE source.  The solid line represents pure photoionization models with ionization parameter $U$ varying between $\log U = -1$ (bottom) and $\log U = -4$ (top), and a spectral index of the ionizing continuum $\alpha = -1$. The dashed line represents a model of shocks and a matter bound precursor, where the shock has $v=1000$ km s$^{-1}$ (models adapted from \citealt{MO02.1}).}
\label{fig:OIHa_HeIIHb}
\end{figure}

To probe the relative roles of photoionization and shocks in the galaxy, we measure the line flux ratios \oiha and \heiihb at each position along the slit where the signal-to-noise ratio is $>3$ for each of the four emission lines.  These line flux ratios discriminate between AGN photoionization and shock models (e.g., \citealt{CL97.1, MO02.1}), and we have illustrated these models in Figure~\ref{fig:OIHa_HeIIHb}.  The photoionization models are computed with {\tt CLOUDY} \citep{FE96.1} and the shock models are computed with {\tt MAPPINGSIII} \citep{DO96.1}.

\begin{figure}[!h]
\begin{center}
\vspace{.2in}
\includegraphics[width=3.2in]{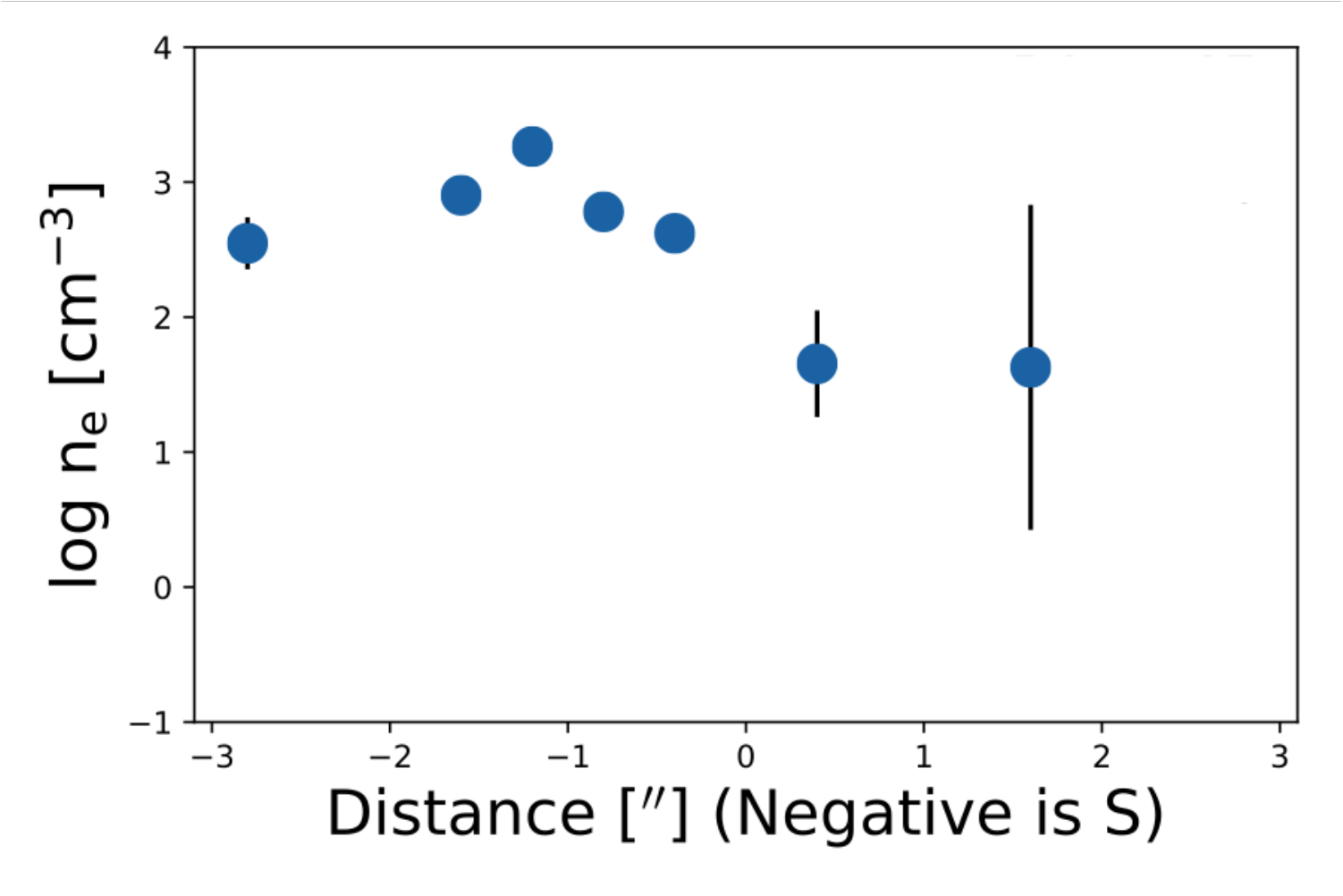}
\end{center}
\vspace{.16in}
\caption{Electron density measured along the APO/DIS longslit spectrum taken at position angle $45^\circ$ east of north. Zero is defined as the position of the host galaxy continuum on the slit, positive spatial positions are to the north, and negative spatial positions are to the south.  Along this position angle, the NE source is located $0\farcs41$ to the north (this position corresponds with the first data point north of the continuum), and the electron density is an order of magnitude lower at the NE source than at the southern positions on the slit.}
\label{fig:ne} 
\end{figure}

In order to explore the densities of the material within the galaxy, we also measure the line flux ratio \siisii at each position along the slit where the signal-to-noise ratio is $>3$ for both emission lines.  Then, we convert this to an electron density \citep{OS06.1,SC16.1} as shown in Figure~\ref{fig:ne}.  We find that the electron densities are consistent with those of a typical narrow-line region ($10^2 \simlt n_e$ (cm$^{-3}$) $\simlt 10^3$; \citealt{OS06.1}), and that the electron density is higher south of the galaxy center and lower north of the galaxy center. 
 
From the longslit spectra, we identify three main kinematic components in the galaxy:

{\it Rotating disk}.  The rotating disk is most apparent at position angle 45 (Figure~\ref{fig:vel}, middle), where the velocities progress from redshifts to the south to blueshifts to the north.  We find that the H$\alpha$ emission is best fit by a rotating disk that has a position angle on the sky of 53$^\circ$ East of North, is inclined 50$^\circ$ into the plane of the sky, has an intrinsic velocity of 182 km s$^{-1}$, has a radius of 2.5 kpc, and whose kinematic center is located at the galaxy's stellar centroid.  This rotating disk is also seen in Pa$\alpha$, with the same disk parameters. 

{\it Spatially extended ionized gas to the south}.  South of the galaxy center, we find ionized gas that extends to distances $\gtrsim4$ kpc from the galaxy center.  We focus our analysis on \oiiiw because it is the strongest emission line and because H$\alpha$ is contaminated by the disk rotation.  We find that the \oiiiw velocities south of the galaxy center range from 24 km s$^{-1}$ redshifted relative to systemic (along position angle 153) to 20 km s$^{-1}$ blueshifted relative to systemic (along position angle 153; Figure~\ref{fig:vel}). The line flux ratios indicate a photoionized origin for emission south of the galaxy center (Figure~\ref{fig:OIHa_HeIIHb}). 

{\it Northern bubble.}  North of the galaxy center there is ionized gas that extends $\sim1$ kpc from the galaxy center.  The \oiiiw velocities north of the galaxy center range from 72 km s$^{-1}$ redshifted relative to systemic (along position angle 45) to 20 km s$^{-1}$ blueshifted relative to systemic (along position angle 0; Figure~\ref{fig:vel}).  The \oiiiw emission is strongest along position angle 45, which is the position angle aligned with the brightest component of the northern bubble seen in the {\it HST} observations. The line flux ratios at the position of the NE source are inconsistent with pure photoionization models, and the high \heiihb ratio can instead be explained by a shock with a precursor \ion{H}{2} region (e.g., \citealt{SU93.1}); \cite{MO02.1} find regions of AGN emission with similar line flux ratios and interpret them also as produced by a shock with a precursor.

\begin{figure*}[!t]
\begin{center}
\includegraphics[width=2.1in]{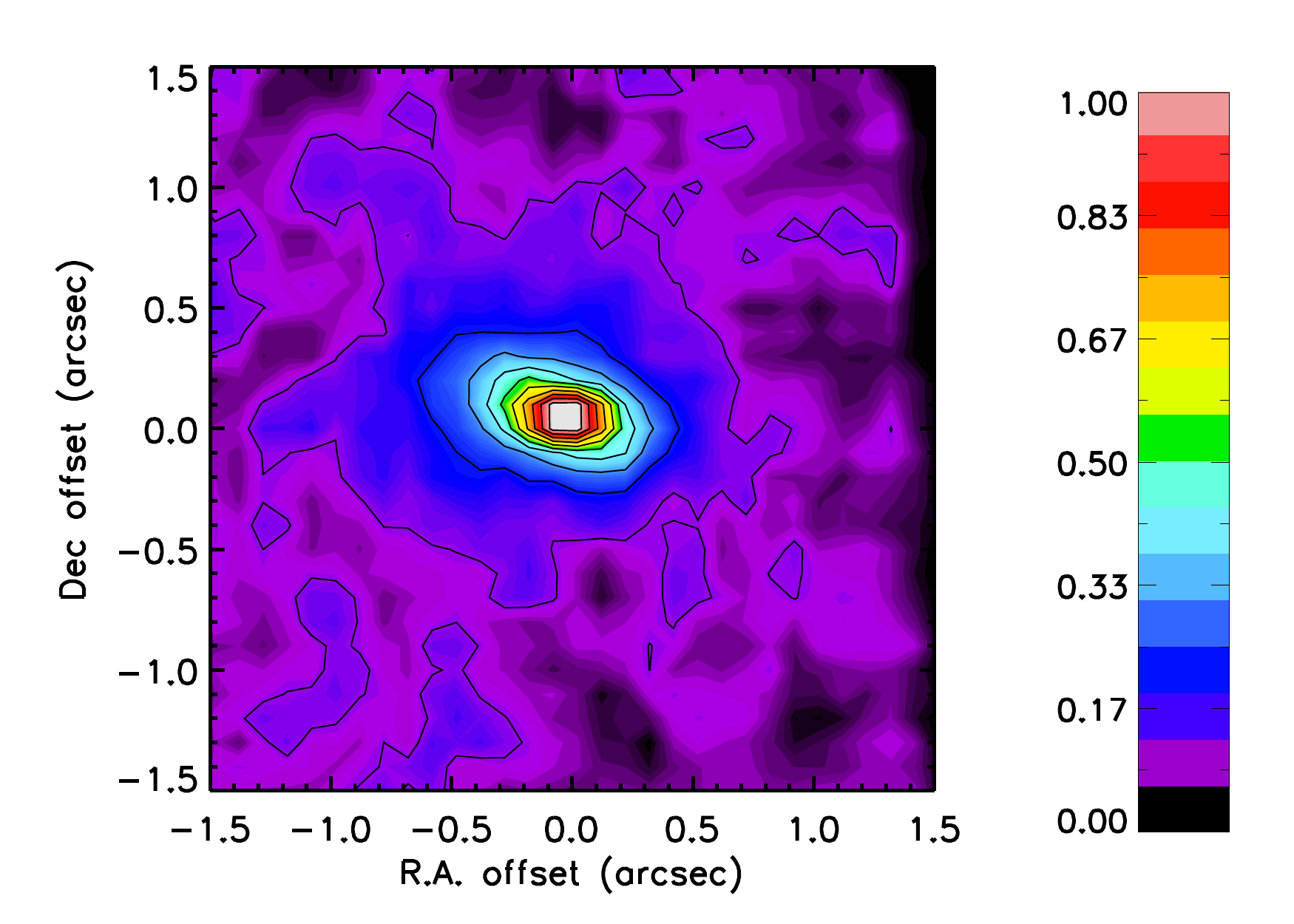}
\includegraphics[width=2.1in]{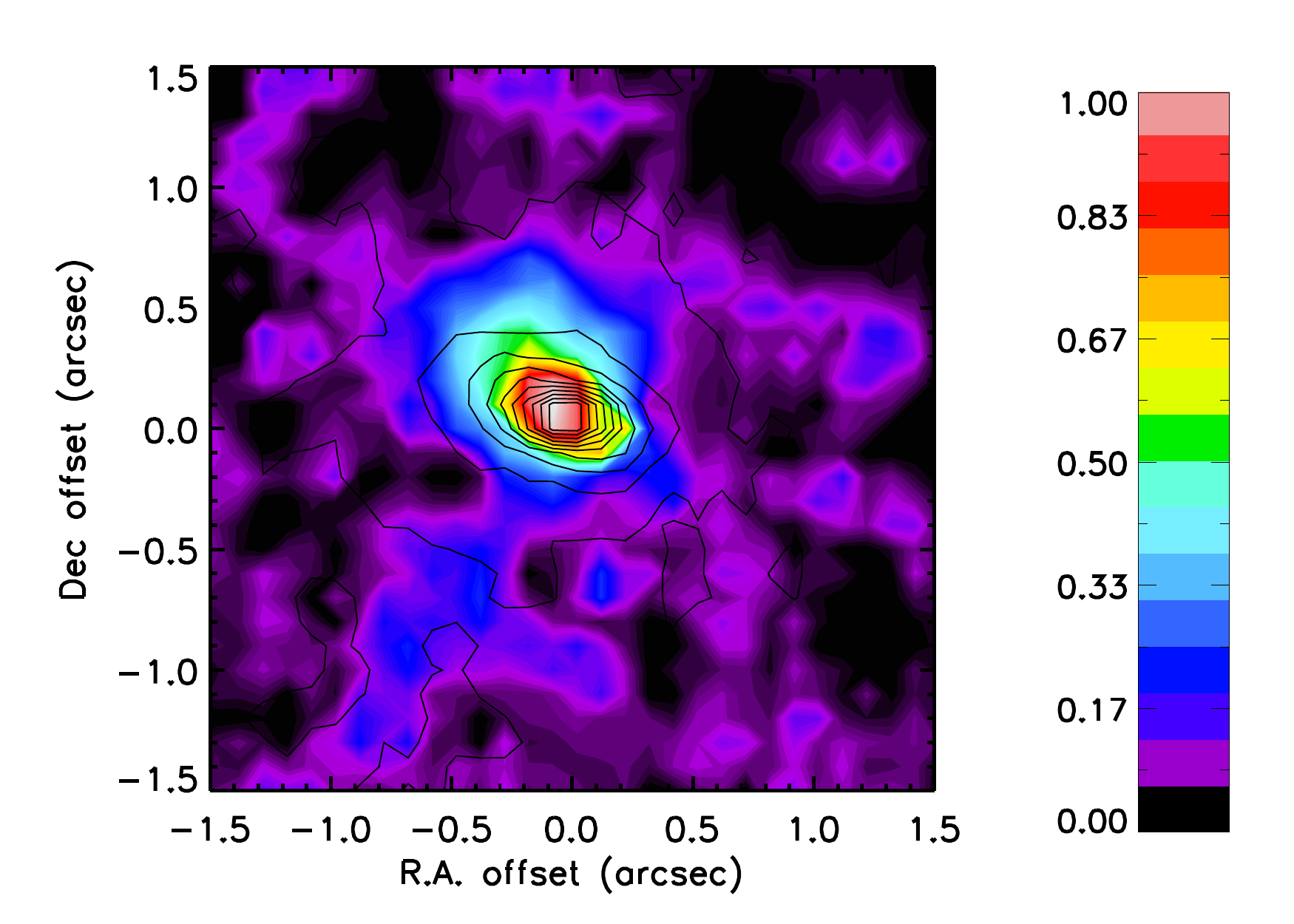}
\includegraphics[width=2.1in]{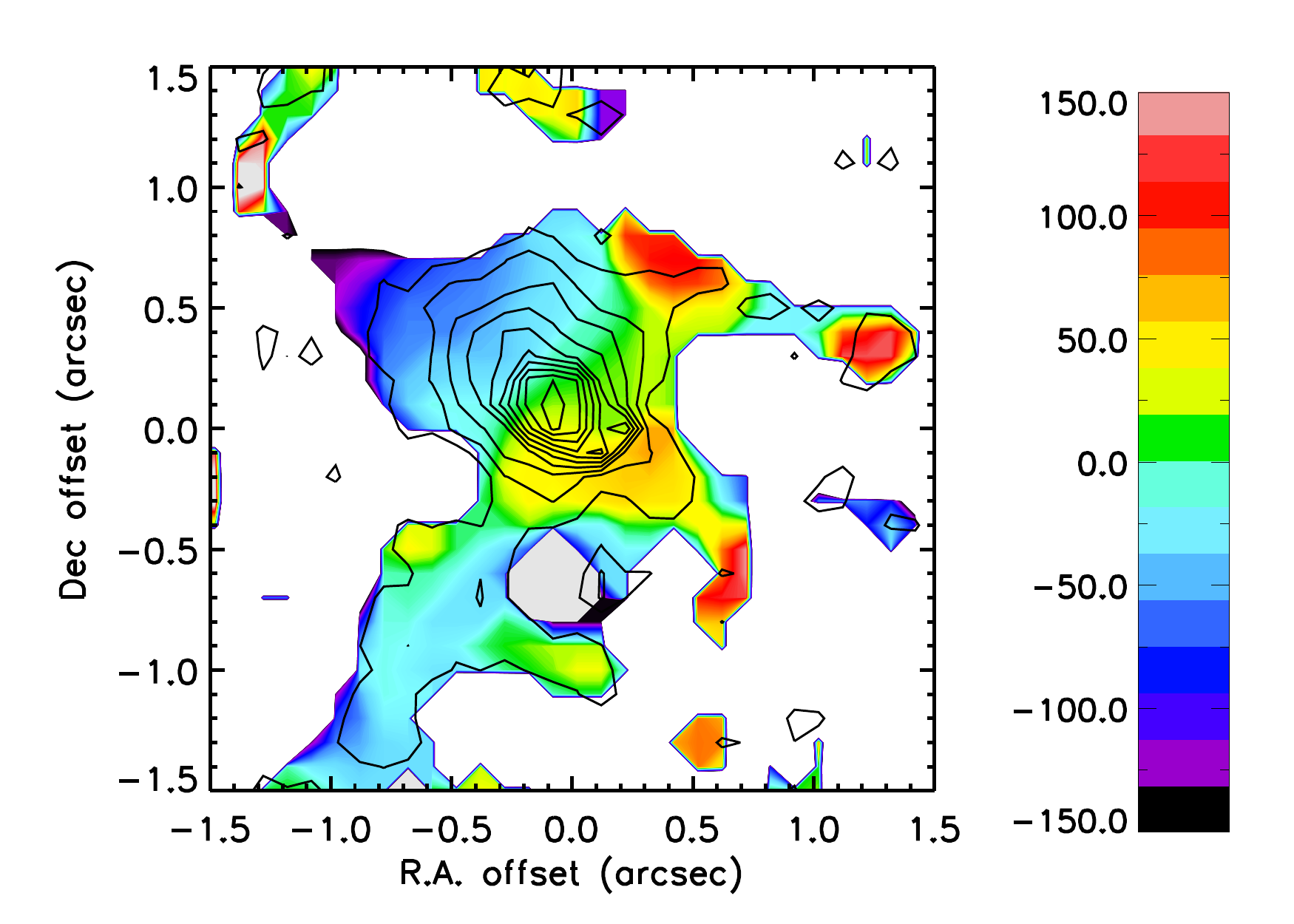}
\includegraphics[width=6.5in]{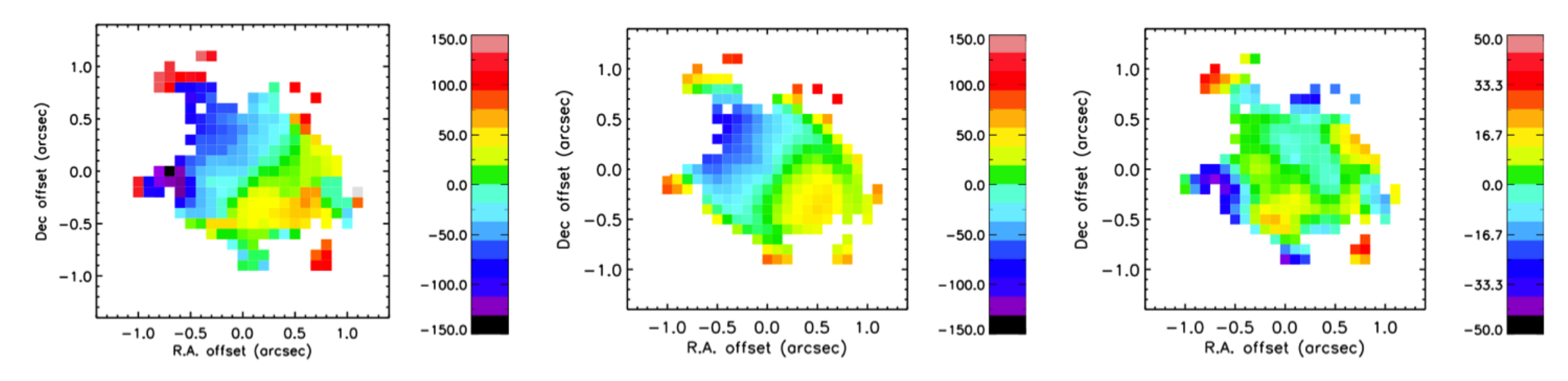}
\end{center}
\caption{Two-dimensional maps of the {\it K}-band continuum (top left), Pa$\alpha$ flux distribution (top middle), and Pa$\alpha$  velocity (top right) for SDSS J1354+1327, where the data have been smoothed with a median filter of 3 pixels. The contours delineate the {\it K}-band continuum emission in the top left and middle panels, and the ionized gas morphology in the top right panel. The center of the galaxy (the peak of continuum emission at 1.9 $\mu$m) is located at position (0,0). Regions in white in the top right panel correspond to pixels where the line properties are uncertain and thus were masked out. These rejected pixels are those with a flux density lower than 5\% of the peak of Pa$\alpha$ emission. Bottom row, from left to right: Pa$\alpha$ velocity map (where pixels that have a flux density lower than 10\% of the peak of the continuum emission have been masked), velocity map of the rotating disk that is the best fit to the data, and the velocity residuals (note the different scale for the velocity color bar).  Each map is centered on the kinematic center of the galaxy.  In all maps north is up and east is to the left.}
\label{fig:osiris_maps}
\end{figure*}

\subsection{Keck/OSIRIS Near-infrared IFS Observations}
\label{osiris}

We observed SDSS J1354+1327 on UT 2013 April 3 using Laser Guide Star Adaptive Optics (LGS-AO) with Keck/OSIRIS integral-field spectroscopy.  We used the Kn1 filter with the $0\farcs1$ pixel scale and the associated $3\farcs6 \times 6\farcs4$ field of view to ensure that the observations would cover the entire area of a $3^{\prime\prime}$ diameter SDSS fiber.  The observations covered an observed wavelength range of 1.955 to 2.055 $\mu$m and detected the Pa$\alpha$ 1.87 $\mu$m emission line, which traces the spatial distribution and kinematics of the ionized gas in the galaxy.  We used the galaxy nucleus as the tip-tilt star, observed at a position angle of 0 degrees East of North, and integrated for 30 minutes.  The data were reduced with the OSIRIS data reduction pipeline, and we used the IDL code LINEFIT \citep{DA07.2} to create 2D kinematic and flux distribution maps (Figure~\ref{fig:osiris_maps}).

From the OSIRIS observations, we identify three main kinematic components in the galaxy:

{\it Rotating disk}.  To model the rotating disk, we first mask the pixels in the Pa$\alpha$ velocity map that have a flux density lower than 10\% of the peak of the continuum emission.  We then use the {\tt KINEMETRY} code \citep{KR06.2} to model the rotating disk structure seen extending from the northeast (blueshifted) to the southwest (redshifted) in Pa$\alpha$. The parameters of the best-fit model are a position angle of 55 degrees East of North, an inclination of 53 degrees, and a kinematic center that is spatially offset by $0\farcs2$ northeast of the peak of the stellar continuum.  The parameters of the rotating disk observed in Pa$\alpha$ are consistent with those of the rotating disk observed in the optical longslit spectra.  After the disk model is subtracted from the data, we examine the velocity residuals, which have $\pm 20$ km s$^{-1}$ errors (Figure~\ref{fig:osiris_maps}, bottom right).  The velocity residuals are not significant enough to fit a secondary component. We find that the kinematic center of the rotating disk is offset by $0\farcs2$ (0.3 kpc) from the stellar continuum center, and we attribute this perturbation to a recent pericenter passage and interaction with the companion galaxy.

{\it Spatially extended ionized gas to the south}.  While the OSIRIS field of view is dominated by the rotating disk, there is also extraplanar Pa$\alpha$ emission visible to the south of the galaxy center, and this emission is detected at $5\sigma$ significance out to the edge of the field of view (Figure~\ref{fig:osiris_maps}, top middle).  The line-of-sight velocities of the gas are $\pm 30$ km s$^{-1}$ .  The Pa$\alpha$ emission also follows the conical structure seen in the {\it HST} observations. 

{\it Northern bubble}.  There is spatially extended Pa$\alpha$ emission associated with the loop seen in the {\it HST} data. 

\section{Results}

\subsection{Nature of the X-ray Source: A Central AGN}
\label{xray-source}

The X-ray source in SDSS J1354+1327 has an observed luminosity of $L_{X, 2-10 \mathrm{keV}}=1.1 \times 10^{43}$ erg s$^{-1}$, which is brighter than the typical AGN luminosity threshold of $L_{X, 2-10 \mathrm{keV}}>10^{42}$ erg s$^{-1}$ and shows that the X-ray source is associated with an AGN.  Mid-infrared colors from the {\it Wide-field Infrared Survey Explorer} ({\it WISE}) also show that an AGN is present (SDSS J1354+1327 has $W1-W2=1.14 \pm 0.03$, where $W1-W2 \geq 0.8$ indicates that the mid-infrared flux is dominated by an AGN; \citealt{ST12.2}).

The fit to the X-ray spectrum reveals a relatively high column density $n_{H,exgal}=2 \times 10^{23}$ cm$^{-2}$, which indicates that the source is an X-ray absorbed AGN.  This is consistent with the $V-H$ dust map, which shows that the very center of the galaxy is the most obscured (Figure~\ref{fig:1354companion}, right).  

Next, we consider the position of the X-ray source within the host galaxy.  We have identified the position of the X-ray source, the positions of the central stellar nucleus of the galaxy and the NE emission source visible in the {\it HST} F438W and F606W observations, and performed astrometric corrections.  We compare the positions to determine if the X-ray AGN position aligns with any of these galaxy features (Figure~\ref{fig:positions}, right).  If the X-ray AGN is associated with the NE source, then the AGN is located 0.40 kpc from the center of SDSS J1354+1327 and this spatial offset could be explained as an offset AGN, a gravitational recoil AGN, or a gravitational slingshot AGN.   We find that the position of the X-ray AGN is more consistent with the position of the stellar centroid of the galaxy (consistent to within $\lesssim1\sigma$) than with the position of the NE source (consistent to within $\lesssim2.5\sigma$).  Consequently, we conclude that the X-ray AGN is most likely located at the galaxy center.  

\subsection{Nature of the Spatially Extended Ionized Gas to the South: A Photoionized AGN Outflow}
\label{outflow}

There is extraplanar ionized gas extending south of the galaxy center, as seen in the {\it HST} F606W and F438W observations, the APO/DIS optical longslit observations, and the Keck/OSIRIS near-infrared IFS observations.  The gas is clumpy, as seen morphologically in the {\it HST} data (Figure~\ref{fig:1354companion}, left) and in the mixture of redshifted and blueshifted line-of-sight velocities of the emission lines observed with APO/DIS (Figure~\ref{fig:vel}).

Here, we aim to determine whether the gas is inflowing, outflowing, or passively photoionized by the central AGN. If the AGN is passively photoionizing the gas in the galaxy, then there should be symmetric photoionized cones on either side of the AGN due to the collimating torus (e.g., \citealt{SC03.3}), but this is not the case for SDSS J1354+1327.
In the passive photoionization scenario, the gas also should be rotating (with possible deviations due to the interaction with the companion galaxy) since it is following the galaxy potential.  We fit a rotating disk to the optical and near-infrared spectra, and we find that the rotating disk is oriented with a position angle 54 degrees East of North, but no rotation is seen in the gas extending to the south of that disk.  Instead the southern gas has blueshifts and redshifts observed along directions not coincident with the axis of rotation, as can be seen in the velocities in the longslit data.  This is evidence of an inflow or outflow, rather than passively photoionized gas.

The morphology of the southern gas distinguishes between the inflow and outflow scenarios.  As shown in the {\it HST} images (Figure~\ref{fig:1354companion}, left), the gas has a conical morphology with the cone beginning at the galaxy center and extending to the south, outside the plane of rotation of the galaxy.  This conical morphology is typical of AGN outflows (e.g., \citealt{MU96.1,SC03.3}), where the torus provides the collimation (e.g., \citealt{AN85.1,MA98.1}).  In contrast, conical morphologies are not expected for inflows, which are typically radial streamers of gas (e.g., \citealt{IO04.1,MU09.1}). 

We conclude that the spatially extended ionized gas to the south is an outflow.  We apply an analytic Markov Chain Monte Carlo model to the longslit observations to model the data as a biconical outflow, as done for similar longslit observations of galaxies in \cite{NE17.1}.  We find that the data are best fit (with a reduced $\chi^2=1.5$) by a bicone extending south of the galaxy center at an inclination of $27^{+14}_{-22}$ degrees out of the plane of the sky, a position angle of $190^{+26}_{-25}$ degrees east of north on the plane of the sky, a half opening angle of $37^{+10}_{-15}$ degrees, a turnover radius of $1.9^{+2.4}_{-1.7}$ kpc, a maximum velocity at the turnover radius $V_{max}=40^{+40}_{-30}$ km s$^{-1}$, and a lateral surface area $A=4^{+14}_{-4}$ kpc$^2$.  The observed blueshifted and redshifted velocities of the gas imply that we are seeing parts of the front and rear facing walls of the outflow (e.g., \citealt{BA16.2}), which fits with the clumpiness of the gas and dust partially obscuring the view (Figure~\ref{fig:1354companion}, right).  When we project the cone onto the plane of the sky, we find an observed opening angle of $66^{+18}_{-27}$ degrees, which is consistent with the opening angle of $67$ degrees measured from the {\it HST} data.

Then, we use the best-fit parameters of the outflow model, our measurement of the electron density $n_e$, and assume a filling factor $f=0.01$ to be conservative in our energy calculation to estimate the mass outflow rate $\dot{M}=m_p n_e V_{max} f A$, where $m_p$ is the proton mass.  We find $\dot{M}=6.5^{+52.3}_{-6.3} \; M_\odot \; \mathrm{yr}^{-1}$.

This outflow could be driven by either star formation or an AGN.  First we consider the scenario of a star formation driven outflow.  
To set an upper limit on the star formation rate (SFR) in SDSS J1354+1327, we assume that all of the \ha emission is associated with star formation.  Using the luminosity measured from the SDSS spectrum, $L_{\han}=1.4 \times 10^{41}$ erg s$^{-1}$, the SFR from \cite{KE12.1} is then $SFR= 5.37 \times 10^{-42} \; L_{\han} = 0.75 \; M_\odot \; \mathrm{yr}^{-1}$.  When we convert this SFR to a mass outflow rate we find an upper limit of $\dot{M}_{SF}=0.2 \; M_\odot \; \mathrm{yr}^{-1}$ \citep{VE05.1}, which is too weak to drive the observed outflow.  We also note that a star formation driven outflow that is powered by supernovae in the disk of the galaxy typically has a wide, chimney-shaped morphology of the gas being driven from the disk into the halo (e.g., \citealt{NO89.1}), which is inconsistent with the conical morphology of the ionized gas observed in SDSS J1354+1327.  

Consequently, we conclude that the outflow is AGN driven.  This is consistent with the \oiiihb and \niiha line flux ratios at the location of the outflow, which show that the emission is Seyfert driven. Further, the \oiha and \heiihb line flux ratios and the conical morphology each indicate that the emission is photoionized (e.g., \citealt{WI94.1}). The AGN luminosity is also sufficient for photoionizing the cone of gas, based on photoionization models computed using the spectral synthesis code \texttt{CLOUDY} \citep{FE13.1, RI14.2}.  We conclude that the spatially extended ionized gas to the south is a photoionized AGN outflow. 

Although we do not know the velocity of the outflow at the time it was launched, we can use the current observed velocity and spatial extent of the outflow to find an upper limit of $10^8$ yrs on the timescale for the outflow to produce the cone of gas.  The light-travel timescale plus the recombination timescale ($t_{rec} \approx (\alpha n_e)^{-1}$, where the recombination coefficient $\alpha=1.72 \times 10^{-11}$ cm$^3$ s$^{-1}$ for \oiii and $n_e$ is from Section~\ref{apo}; \citealt{OS06.1}) to illuminate the gas is then $\sim50$ yrs, which is negligible.  Given that the emission can be observed for up to $10^5$ yrs (e.g., \citealt{SC15.1}), we conclude that the outflow was likely launched $\simlt10^5$ yrs ago.

\begin{figure*}[!t]
\begin{center}
\includegraphics[width=7.2in]{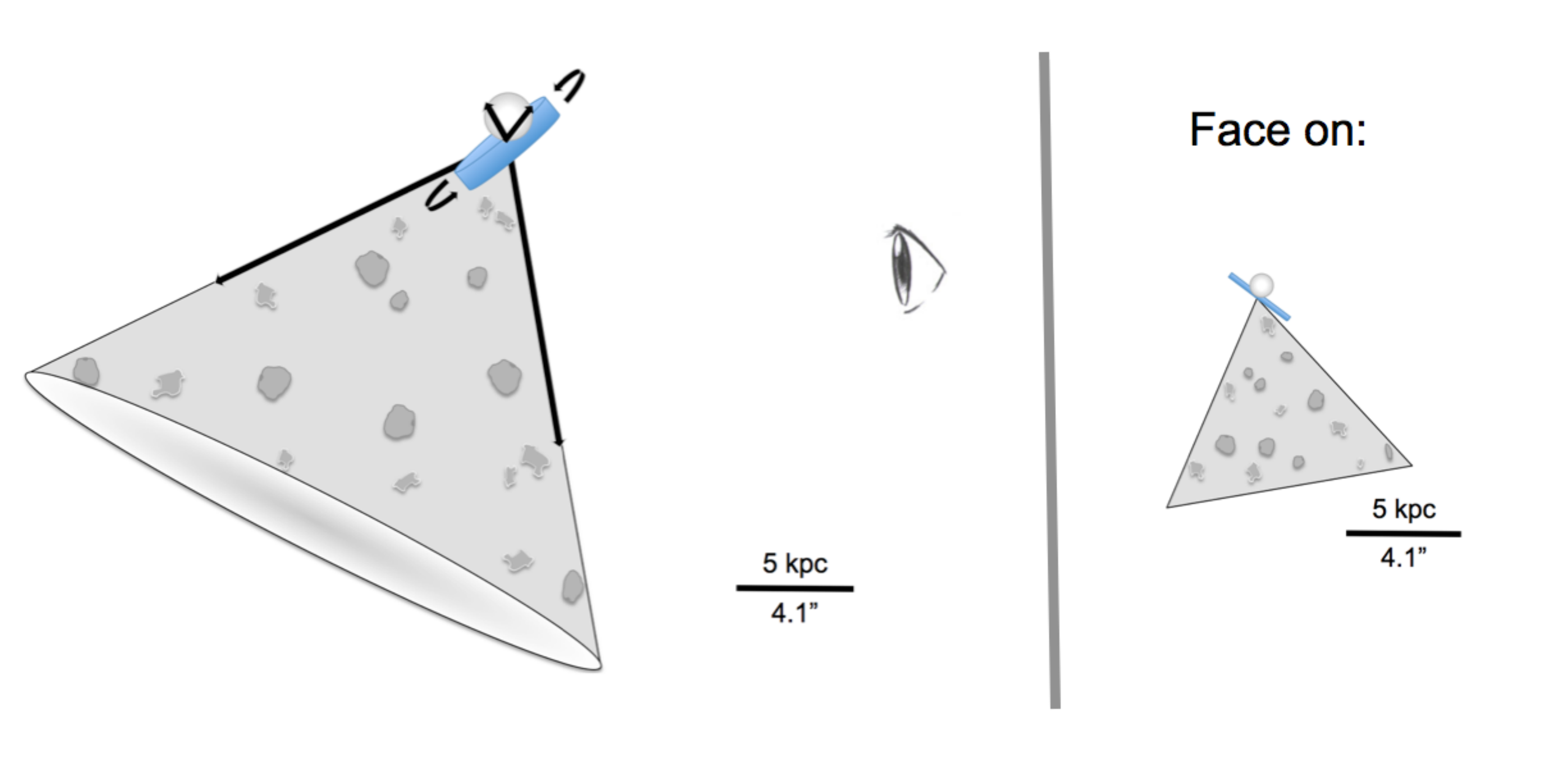}
\end{center}
\caption{Left: 3D drawing of the main components of SDSS J1354+1327: a southern conical outflow (with clumpiness shown), a northern bubble, and a rotating disk.  Arrows show the directions of the gas motion in each component, and an eye shows the orientation of the observer. Right: face on view of the galaxy's main components, as shown from the perspective of the observer.}
\label{fig:drawing}
\end{figure*}

\subsection{Nature of the Northern Bubble: A Shocked AGN Outflow}
\label{ne}

The {\it HST} F438W and F606W observations reveal a loop of gas extending 0.7 kpc north of the galaxy center, as well as a NE source that is located 0.40 kpc northeast of the galaxy center and is embedded in the loop of emission (Figure~\ref{fig:1354companion}, left). The peak of \oiiiw emission is located at the NE source as well.

As we did with the spatially extended ionized gas to the south, we consider whether the northern bubble is caused by gas that is inflowing, outflowing, or passively photoionized by the AGN.  First, the off-nuclear peak of the \oiiiw emission indicates a shock at the position of the NE source and explains the velocity-offset emission line in the SDSS spectrum.  We also find that the observed line flux ratios cannot be explained by pure photoionization, and instead are explained by shocks. Further evidence for shocks comes from the looped morphology of the northern bubble, which indicates that shocks are being driven into the ambient gas (e.g., \citealt{FA12.3,GA14.2}).  Where we see redshifted velocities, we are seeing the back side of the bubble.

The presence of shocks rules out the scenario of an AGN passively photoionizing the northern gas in SDSS J1354+1327, and leaves inflowing or outflowing gas as the remaining scenarios.  Inflowing gas typically streams radially towards the galaxy center (e.g., \citealt{IO04.1, MU09.1}), and the bubble morphology of the observed gas is difficult to explain for an inflow.  In contrast, many AGN outflows are seen blowing bubbles of gas (e.g., \citealt{FA12.2,GR14.1}).  We conclude that the northern bubble is an AGN outflow driving shocks into the ambient material.

The original velocity of the outflow is unknown, but we have measured the current observed velocity and spatial extent of the bubble.  Using these values, we estimate that the outflow was launched $\simlt10^7$ yrs ago.  Taking into account the length of time that the emission signatures can linger, the outflow must have been launched $\simlt10^5$ yrs ago.

\subsection{Companion Galaxy to SDSS J1354+1327}
\label{companion}

SDSS J1354+1327 has a companion galaxy, SDSS J135429.17+132807.3, that is located 12.5 kpc to the northeast and is redshifted by 76 km s$^{-1}$ relative to SDSS J1354+1327. Emission line diagnostics of the companion's SDSS spectrum suggest that it hosts an AGN, and \cite{LI11.3} also noted that SDSS J1354+1327 and SDSS J1354+1328 are a pair of active galaxies.  However, there is no {\it WISE} source detected at the location of the companion galaxy.  Further, we do not detect an AGN in our {\it Chandra} observations of the companion galaxy.  

The lack of an X-ray detection suggests that the AGN is obscured, the AGN's X-ray luminosity is low on the predicted \oiiiw to X-ray luminosity scaling relation, some of the gas covered by the SDSS fiber is ionized by SDSS J1354+1327's AGN, or that the \oiiiw emission is a light echo of past AGN activity (e.g., \citealt{LI09.2,SC15.1}; the AGN may be flickering on and off, or the SMBH may have been ejected from the host galaxy center).  The other possibility is that the \oiiiw luminosity is not associated with an AGN.  In this scenario, the observed line flux ratio ($\log$ \oiiihb$=0.44$) could be produced by a starburst instead of an AGN, given that models predict that starbursts can produce line ratios that high. Given that the dust map indicates obscuration at the galaxy center ($V-H=2.4$), we conclude that the galaxy most likely hosts an obscured AGN.

The {\it HST} observations show tidal tails of stars connecting the primary and companion galaxies, indicating that they are already interacting (Figure~\ref{fig:positions}, bottom left).  Taking the mass ratio to be the luminosity ratio of the two stellar bulges, the merger's mass ratio is $0.98$ (SDSS J1354+1327 is slightly less massive).  

\section{Nature of SDSS J1354+1327: Episodic AGN Outbursts}
\label{nature}

We have shown that SDSS J1354+1327 hosts asymmetric AGN-driven outflows.  South of the galaxy center there is a large, $\sim10$ kpc extended outflow of ionized gas, while north of the galaxy center there is a small scale ($\sim1$ kpc) bubble from a shocked AGN outflow.  The main components of SDSS J1354+1327 are illustrated in a diagram in Figure~\ref{fig:drawing}.

One possible explanation for the asymmetry of the outflow is that there were concurrent AGN outbursts to the north and south, but the southern outburst traveled more efficiently while the northern outburst was blocked by high density ISM clouds (e.g., \citealt{DE17.1}).  In this scenario, dense clouds to the north of the AGN would absorb the outflow energy and radiate it away, while the southern outflow would propagate unimpeded.  However, in SDSS J1354+1327 we find that the density at the northern outburst ($n_e=45$ cm$^{-3}$) is an order of magnitude lower than the densities associated with the southern outburst ($n_e \geq 420$ cm$^{-3}$), which does not support the concurrent AGN outburst scenario.

Another possible explanation for the asymmetric outflows in SDSS J1354+1327 is discontinuous AGN outburst activity, which has been seen in simulations and can create outbursts on opposite sides of the galaxy such as those seen in SDSS J1354+1327 \citep{HO10.1,GA14.2}.  Clumpy structure in the ISM can lead to highly variable accretion rates onto the supermassive black hole (e.g., \citealt{HI14.1}), where each high-accretion event produces an AGN outburst.  This outburst injects thermal energy into the ambient gas, launching an outflow.  Initially, the heated gas encounters and shocks cold gas, making shocks an indicator of a recent AGN outburst \citep{GA14.2}.  After this initial  period of fast expansion and shocks, the outflow slows and photoionization of the surrounding gas becomes the dominant emission source.  Consequently, two sequential AGN outbursts can produce asymmetric outflows; the first AGN outburst leaves behind a large photoionized cone of emission, while the later AGN outburst produces small-scale shock emission.

The observations support this picture of two sequential AGN outbursts.  We interpret the southern outflow as the result of an older AGN outburst, based on the photoionized nature of the gas (indicated by the optical line flux ratios and the gas's conical morphology) and the large spatial extent ($\sim10$ kpc) of the gas.  We interpret the northern outflow as a new outflow that is being driven by a recent AGN outburst, based on the shocked nature of the gas (indicated by the optical line flux ratios, the off-nuclear peak in \oiiiwn, and the gas's looped morphology) and the small spatial extent ($\sim1$ kpc) of the gas.

Therefore, the likeliest explanation for the nature of SDSS J1354+1327 is that it experienced two sequential AGN outbursts.  These episodic AGN outbursts could be the result of a galaxy merger that drives discrete accretion events onto the supermassive black hole that are followed by discrete AGN outbursts (e.g., \citealt{GA13.1}).  SDSS J1354+1327 has a companion galaxy located 12.5 kpc to the northeast, and the the tidal debris connecting the two galaxies and the spatial offset between the kinematic center of the galaxy and the center of the stellar light suggest that the two galaxies have already had a close pericenter passage.  The interaction with the companion galaxy could have triggered nuclear inflows of gas in SDSS J1354+1327, which produced the two episodes of accretion onto the SMBH that led to the two consecutive AGN outbursts.

There have been many other observations of asymmetric AGN outflows (e.g., \citealt{SH98.1,FI11.1,GR12.2}), but the origins of these asymmetries have not yet been pinpointed.  The variety of sizes and velocity profiles observed in AGN outflows could be explained by the lifecycle of AGN flickering \citep{SU16.1}.  There also exist AGN outflows with little to no indication of present AGN activity (e.g., \citealt{TR07.1,FO14.1}) and many examples of AGN light echoes (e.g., \citealt{LI09.2,KE15.1,RU16.1}), which may signal SMBHs transitioning from the luminous AGN phase to quiescence.

The galaxy that is the most similar to SDSS J1354+1327 is the $z=0.12$ galaxy SDSS J1356+1026.  SDSS J1356+1026 also has velocity-offset emission lines in its SDSS spectrum (although they are double peaked, instead of single peaked as for SDSS J1354+1327), large-scale AGN-driven outflows, and a companion galaxy located 3 kpc away \citep{LI10.1,FU12.1,GR12.2,GR14.1,CO15.1}.  The major difference is that SDSS J1356+1026 is much more luminous (bolometric luminosity $1.5 \times 10^{46}$ erg s$^{-1}$) and powerful: SDSS J1356+1026 is an ultra-luminous infrared galaxy hosting a quasar that is powering large-scale ($\sim10$ kpc) symmetric $\sim1000$ km s$^{-1}$ outflows of ionized gas, and it also hosts a compact ($\sim0.3$ kpc) structure of molecular gas \citep{GR12.2,SU14.1}.  If the molecular gas is in an outflow, then the outflows in SDSS J1356+1026 may be driven by episodic AGN activity, where the ionized outflows were launched before the molecular outflow \citep{SU14.1}.

We conclude that SDSS J1354+1327 is the galaxy with the strongest evidence for both a light echo of a previous AGN outflow and a new AGN outflow being launched.  This is direct observational evidence for the episodic AGN outbursts that are seen in simulations.  Given the upper limits on when the outflows launched, and the amount of time that emission can remain visible, we infer that the timescale between the two AGN outbursts was $\lesssim10^5$ yrs.

\section{Conclusions}

We have used {\it Chandra}, {\it HST}, APO/DIS optical longslit spectroscopy, and Keck/OSIRIS near-infrared AO-assisted IFS to determine the nature of the $z=0.06$ galaxy SDSS J1354+1327.   SDSS J1354+1327 was originally selected from SDSS spectra of active galaxies for its narrow AGN emission lines that are offset in line-of-sight velocity from systemic by 69 km s$^{-1}$.  We determined that the velocity offset arises from the peak in emission line gas, which is located 0.40 kpc northeast of the galaxy center.

We conclude that SDSS J1354+1327 is a galaxy whose AGN has recently turned on and then on again, through two accretion events that led to two sequential AGN outbursts.  The two accretion events arose from clumpy ISM being driven to the center of SDSS J1354+1327 as a result of a tidal interaction with the companion galaxy located 12.5 kpc away.  The outflow visible to the south of the galaxy center is remnant emission lingering from a past AGN outburst, because it is photoionized gas with a large spatial extent of the gas ($\sim 10$ kpc).  The emission north of the galaxy center is a new AGN outburst that launched a new outflow, based on the shocked nature of the gas and the smaller spatial extent of the gas ($\sim1$ kpc).  We estimate that the timescale between the AGN outburst that produces the southern outflow and the AGN outburst that is powering the northern outflow is $\lesssim10^5$ yrs. 

SDSS J1354+1327 is the strongest observational example of a galaxy with an AGN that has turned off and then on again.  As such, SDSS J1354+1327 fits into the broader picture of episodic AGN activity that also includes many observations of light echoes of past AGN outbursts.  Our attention was initially drawn to SDSS J1354+1327 due to its narrow emission lines that are offset in line-of-sight velocity from systemic.  These velocity-offset emission lines are commonly indicators of shocks, and one mechanism that produces shocks is a recent AGN outburst being driven into the ISM.  Therefore, future analyses of galaxies with velocity-offset AGN emission lines will certainly uncover young AGN outbursts and some fraction of them may have evidence of past AGN activity as well.

\acknowledgements J.M.C. thanks Nadia Zakamska for useful discussions.  We also thank the anonymous referee for comments that have improved the clarity of this paper.  Support for this work was provided by NASA through Chandra Award Number GO4-15113X issued by the Chandra X-ray Observatory Center, which is operated by the Smithsonian Astrophysical Observatory for and on behalf of NASA under contract NAS8-03060.  Support for HST program number GO-13513 was provided by NASA through a grant from the Space Telescope Science Institute, which is operated by the Association of Universities for Research in Astronomy, Inc., under NASA contract NAS5-26555.

The scientific results reported in this article are based in part on observations made by the Chandra X-ray Observatory, and this research has made use of software provided by the Chandra X-ray Center in the application packages CIAO, ChIPS, and Sherpa.  The results reported here are also based on observations made with the NASA/ESA Hubble Space Telescope, obtained at the Space Telescope Science Institute, which is operated by the Association of Universities for Research in Astronomy, Inc., under NASA contract NAS 5-26555. These observations are associated with program number GO-13513.

Some of the observations reported here were obtained at the Apache Point Observatory 3.5m telescope, which is owned and operated by the Astrophysical Research Consortium.  

Some of the data presented herein were obtained at the W.M. Keck Observatory, which is operated as a scientific partnership among the California Institute of Technology, the University of California and the National Aeronautics and Space Administration. The Observatory was made possible by the generous financial support of the W.M. Keck Foundation. The authors wish to recognize and acknowledge the very significant cultural role and reverence that the summit of Mauna Kea has always had within the indigenous Hawaiian community. We are most fortunate to have the opportunity to conduct observations from this mountain.

{\it Facilities:} \facility{{\it CXO} (ACIS)}, \facility{{\it HST} (WFC3)}, \facility{APO (Dual Imaging spectrograph)}, \facility{Keck:I (OSIRIS)}

\bibliographystyle{apj}

\begin{thebibliography}{84}
\expandafter\ifx\csname natexlab\endcsname\relax\def\natexlab#1{#1}\fi

\bibitem[{{Allen} {et~al.}(2015){Allen}, {Schaefer}, {Scott}, {Fogarty}, {Ho},
  {Medling}, {Leslie}, {Bland-Hawthorn}, {Bryant}, {Croom}, {Goodwin}, {Green},
  {Konstantopoulos}, {Lawrence}, {Owers}, {Richards}, \& {Sharp}}]{AL15.1}
{Allen}, J.~T., {Schaefer}, A.~L., {Scott}, N., {Fogarty}, L.~M.~R., {Ho},
  I.-T., {Medling}, A.~M., {Leslie}, S.~K., {Bland-Hawthorn}, J., {Bryant},
  J.~J., {Croom}, S.~M., {Goodwin}, M., {Green}, A.~W., {Konstantopoulos},
  I.~S., {Lawrence}, J.~S., {Owers}, M.~S., {Richards}, S.~N., \& {Sharp}, R.
  2015, \mnras, 451, 2780

\bibitem[{{Antonucci} \& {Miller}(1985)}]{AN85.1}
{Antonucci}, R.~R.~J., \& {Miller}, J.~S. 1985, \apj, 297, 621

\bibitem[{{Bae} \& {Woo}(2016)}]{BA16.2}
{Bae}, H.-J., \& {Woo}, J.-H. 2016, \apj, 828, 97

\bibitem[{{Baldwin} {et~al.}(1981){Baldwin}, {Phillips}, \&
  {Terlevich}}]{BA81.1}
{Baldwin}, J.~A., {Phillips}, M.~M., \& {Terlevich}, R. 1981, \pasp, 93, 5

\bibitem[{{Barrows} {et~al.}(2016){Barrows}, {Comerford}, {Greene}, \&
  {Pooley}}]{BA16.1}
{Barrows}, R.~S., {Comerford}, J.~M., {Greene}, J.~E., \& {Pooley}, D. 2016,
  \apj, 829, 37

\bibitem[{{Barrows} {et~al.}(2013){Barrows}, {Sandberg Lacy}, {Kennefick},
  {Comerford}, {Kennefick}, \& {Berrier}}]{BA13.1}
{Barrows}, R.~S., {Sandberg Lacy}, C.~H., {Kennefick}, J., {Comerford}, J.~M.,
  {Kennefick}, D., \& {Berrier}, J.~C. 2013, \apj, 769, 95

\bibitem[{{Barrows} {et~al.}(2012){Barrows}, {Stern}, {Madsen}, {Harrison},
  {Assef}, {Comerford}, {Cushing}, {Fassnacht}, {Gonzalez}, {Griffith},
  {Hickox}, {Kirkpatrick}, \& {Lagattuta}}]{BA12.1}
{Barrows}, R.~S., {Stern}, D., {Madsen}, K., {Harrison}, F., {Assef}, R.~J.,
  {Comerford}, J.~M., {Cushing}, M.~C., {Fassnacht}, C.~D., {Gonzalez}, A.~H.,
  {Griffith}, R., {Hickox}, R., {Kirkpatrick}, J.~D., \& {Lagattuta}, D.~J.
  2012, \apj, 744, 7

\bibitem[{{Bevington}(1969)}]{BE69.1}
{Bevington}, P.~R. 1969, {Data reduction and error analysis for the physical
  sciences}

\bibitem[{{Clark} {et~al.}(1997){Clark}, {Tadhunter}, {Morganti}, {Killeen},
  {Fosbury}, {Hook}, {Siebert}, \& {Shaw}}]{CL97.1}
{Clark}, N.~E., {Tadhunter}, C.~N., {Morganti}, R., {Killeen}, N.~E.~B.,
  {Fosbury}, R.~A.~E., {Hook}, R.~N., {Siebert}, J., \& {Shaw}, M.~A. 1997,
  \mnras, 286, 558

\bibitem[{{Comerford} {et~al.}(2009){Comerford}, {Gerke}, {Newman}, {Davis},
  {Yan}, {Cooper}, {Faber}, {Koo}, {Coil}, {Rosario}, \& {Dutton}}]{CO09.1}
{Comerford}, J.~M., {Gerke}, B.~F., {Newman}, J.~A., {Davis}, M., {Yan}, R.,
  {Cooper}, M.~C., {Faber}, S.~M., {Koo}, D.~C., {Coil}, A.~L., {Rosario},
  D.~J., \& {Dutton}, A.~A. 2009, \apj, 698, 956

\bibitem[{{Comerford} {et~al.}(2012){Comerford}, {Gerke}, {Stern}, {Cooper},
  {Weiner}, {Newman}, {Madsen}, \& {Barrows}}]{CO12.1}
{Comerford}, J.~M., {Gerke}, B.~F., {Stern}, D., {Cooper}, M.~C., {Weiner},
  B.~J., {Newman}, J.~A., {Madsen}, K., \& {Barrows}, R.~S. 2012, \apj, 753, 42

\bibitem[{{Comerford} \& {Greene}(2014)}]{CO14.1}
{Comerford}, J.~M., \& {Greene}, J.~E. 2014, \apj, 789, 112

\bibitem[{{Comerford} {et~al.}(2015){Comerford}, {Pooley}, {Barrows}, {Greene},
  {Zakamska}, {Madejski}, \& {Cooper}}]{CO15.1}
{Comerford}, J.~M., {Pooley}, D., {Barrows}, R.~S., {Greene}, J.~E.,
  {Zakamska}, N.~L., {Madejski}, G.~M., \& {Cooper}, M.~C. 2015, \apj, 806, 219

\bibitem[{{Comerford} {et~al.}(2011){Comerford}, {Pooley}, {Gerke}, \&
  {Madejski}}]{CO11.2}
{Comerford}, J.~M., {Pooley}, D., {Gerke}, B.~F., \& {Madejski}, G.~M. 2011,
  \apjl, 737, L19+

\bibitem[{{Comerford} {et~al.}(2013){Comerford}, {Schluns}, {Greene}, \&
  {Cool}}]{CO13.1}
{Comerford}, J.~M., {Schluns}, K., {Greene}, J.~E., \& {Cool}, R.~J. 2013,
  \apj, 777, 64

\bibitem[{{Cutri} {et~al.}(2003){Cutri}, {Skrutskie}, {van Dyk}, {Beichman},
  {Carpenter}, {Chester}, {Cambresy}, {Evans}, {Fowler}, {Gizis}, {Howard},
  {Huchra}, {Jarrett}, {Kopan}, {Kirkpatrick}, {Light}, {Marsh}, {McCallon},
  {Schneider}, {Stiening}, {Sykes}, {Weinberg}, {Wheaton}, {Wheelock}, \&
  {Zacarias}}]{CU03.1}
{Cutri}, R.~M., {Skrutskie}, M.~F., {van Dyk}, S., {Beichman}, C.~A.,
  {Carpenter}, J.~M., {Chester}, T., {Cambresy}, L., {Evans}, T., {Fowler}, J.,
  {Gizis}, J., {Howard}, E., {Huchra}, J., {Jarrett}, T., {Kopan}, E.~L.,
  {Kirkpatrick}, J.~D., {Light}, R.~M., {Marsh}, K.~A., {McCallon}, H.,
  {Schneider}, S., {Stiening}, R., {Sykes}, M., {Weinberg}, M., {Wheaton},
  W.~A., {Wheelock}, S., \& {Zacarias}, N. 2003, VizieR Online Data Catalog,
  2246

\bibitem[{{Davies} {et~al.}(2007){Davies}, {Mueller S{\'a}nchez}, {Genzel},
  {Tacconi}, {Hicks}, {Friedrich}, \& {Sternberg}}]{DA07.2}
{Davies}, R.~I., {Mueller S{\'a}nchez}, F., {Genzel}, R., {Tacconi}, L.~J.,
  {Hicks}, E.~K.~S., {Friedrich}, S., \& {Sternberg}, A. 2007, \apj, 671, 1388

\bibitem[{{DeGraf} {et~al.}(2017){DeGraf}, {Dekel}, {Gabor}, \&
  {Bournaud}}]{DE17.1}
{DeGraf}, C., {Dekel}, A., {Gabor}, J., \& {Bournaud}, F. 2017, \mnras, 466,
  1462

\bibitem[{{Dickey} \& {Lockman}(1990)}]{DI90.1}
{Dickey}, J.~M., \& {Lockman}, F.~J. 1990, \araa, 28, 215

\bibitem[{{Dopita} \& {Sutherland}(1996)}]{DO96.1}
{Dopita}, M.~A., \& {Sutherland}, R.~S. 1996, \apjs, 102, 161

\bibitem[{{Fabian}(2012)}]{FA12.2}
{Fabian}, A.~C. 2012, \araa, 50, 455

\bibitem[{{Faucher-Gigu{\`e}re} \& {Quataert}(2012)}]{FA12.3}
{Faucher-Gigu{\`e}re}, C.-A., \& {Quataert}, E. 2012, \mnras, 425, 605

\bibitem[{{Ferland}(1996)}]{FE96.1}
{Ferland}, G.~J. 1996, {Hazy, A Brief Introduction to Cloudy 90}

\bibitem[{{Ferland} {et~al.}(2013){Ferland}, {Porter}, {van Hoof}, {Williams},
  {Abel}, {Lykins}, {Shaw}, {Henney}, \& {Stancil}}]{FE13.1}
{Ferland}, G.~J., {Porter}, R.~L., {van Hoof}, P.~A.~M., {Williams}, R.~J.~R.,
  {Abel}, N.~P., {Lykins}, M.~L., {Shaw}, G., {Henney}, W.~J., \& {Stancil},
  P.~C. 2013, RMXAA, 49, 137

\bibitem[{{Fischer} {et~al.}(2011){Fischer}, {Crenshaw}, {Kraemer}, {Schmitt},
  {Mushotsky}, \& {Dunn}}]{FI11.1}
{Fischer}, T.~C., {Crenshaw}, D.~M., {Kraemer}, S.~B., {Schmitt}, H.~R.,
  {Mushotsky}, R.~F., \& {Dunn}, J.~P. 2011, \apj, 727, 71

\bibitem[{{F{\"o}rster Schreiber} {et~al.}(2014){F{\"o}rster Schreiber},
  {Genzel}, {Newman}, {Kurk}, {Lutz}, {Tacconi}, {Wuyts}, {Bandara}, {Burkert},
  {Buschkamp}, {Carollo}, {Cresci}, {Daddi}, {Davies}, {Eisenhauer}, {Hicks},
  {Lang}, {Lilly}, {Mainieri}, {Mancini}, {Naab}, {Peng}, {Renzini}, {Rosario},
  {Shapiro Griffin}, {Shapley}, {Sternberg}, {Tacchella}, {Vergani},
  {Wisnioski}, {Wuyts}, \& {Zamorani}}]{FO14.1}
{F{\"o}rster Schreiber}, N.~M., {Genzel}, R., {Newman}, S.~F., {Kurk}, J.~D.,
  {Lutz}, D., {Tacconi}, L.~J., {Wuyts}, S., {Bandara}, K., {Burkert}, A.,
  {Buschkamp}, P., {Carollo}, C.~M., {Cresci}, G., {Daddi}, E., {Davies}, R.,
  {Eisenhauer}, F., {Hicks}, E.~K.~S., {Lang}, P., {Lilly}, S.~J., {Mainieri},
  V., {Mancini}, C., {Naab}, T., {Peng}, Y., {Renzini}, A., {Rosario}, D.,
  {Shapiro Griffin}, K., {Shapley}, A.~E., {Sternberg}, A., {Tacchella}, S.,
  {Vergani}, D., {Wisnioski}, E., {Wuyts}, E., \& {Zamorani}, G. 2014, \apj,
  787, 38

\bibitem[{{Fu} {et~al.}(2012){Fu}, {Yan}, {Myers}, {Stockton}, {Djorgovski},
  {Aldering}, \& {Rich}}]{FU12.1}
{Fu}, H., {Yan}, L., {Myers}, A.~D., {Stockton}, A., {Djorgovski}, S.~G.,
  {Aldering}, G., \& {Rich}, J.~A. 2012, \apj, 745, 67

\bibitem[{{Gabor} \& {Bournaud}(2013)}]{GA13.1}
{Gabor}, J.~M., \& {Bournaud}, F. 2013, \mnras, 434, 606

\bibitem[{{Gabor} \& {Bournaud}(2014)}]{GA14.2}
---. 2014, \mnras, 441, 1615

\bibitem[{{Graham} \& {Driver}(2005)}]{GR05.2}
{Graham}, A.~W., \& {Driver}, S.~P. 2005, PASA, 22, 118

\bibitem[{{Greene} {et~al.}(2014){Greene}, {Pooley}, {Zakamska}, {Comerford},
  \& {Sun}}]{GR14.1}
{Greene}, J.~E., {Pooley}, D., {Zakamska}, N.~L., {Comerford}, J.~M., \& {Sun},
  A.-L. 2014, \apj, 788, 54

\bibitem[{{Greene} {et~al.}(2012){Greene}, {Zakamska}, \& {Smith}}]{GR12.2}
{Greene}, J.~E., {Zakamska}, N.~L., \& {Smith}, P.~S. 2012, \apj, 746, 86

\bibitem[{{Hickox} {et~al.}(2014){Hickox}, {Mullaney}, {Alexander}, {Chen},
  {Civano}, {Goulding}, \& {Hainline}}]{HI14.1}
{Hickox}, R.~C., {Mullaney}, J.~R., {Alexander}, D.~M., {Chen}, C.-T.~J.,
  {Civano}, F.~M., {Goulding}, A.~D., \& {Hainline}, K.~N. 2014, \apj, 782, 9

\bibitem[{{Hopkins} \& {Quataert}(2010)}]{HO10.1}
{Hopkins}, P.~F., \& {Quataert}, E. 2010, \mnras, 407, 1529

\bibitem[{{Iono} {et~al.}(2004){Iono}, {Yun}, \& {Mihos}}]{IO04.1}
{Iono}, D., {Yun}, M.~S., \& {Mihos}, J.~C. 2004, \apj, 616, 199

\bibitem[{Ishibashi \& Courvoisier(2010)}]{IS10.1}
Ishibashi, W., \& Courvoisier, T.-L. 2010, Astronomy \& Astrophysics, 512, A58

\bibitem[{{Kauffmann} {et~al.}(2003){Kauffmann}, {Heckman}, {Tremonti},
  {Brinchmann}, {Charlot}, {White}, {Ridgway}, {Brinkmann}, {Fukugita}, {Hall},
  {Ivezi{\'c}}, {Richards}, \& {Schneider}}]{KA03.1}
{Kauffmann}, G., {Heckman}, T.~M., {Tremonti}, C., {Brinchmann}, J., {Charlot},
  S., {White}, S.~D.~M., {Ridgway}, S.~E., {Brinkmann}, J., {Fukugita}, M.,
  {Hall}, P.~B., {Ivezi{\'c}}, {\v Z}., {Richards}, G.~T., \& {Schneider},
  D.~P. 2003, \mnras, 346, 1055

\bibitem[{{Keel} {et~al.}(2015){Keel}, {Maksym}, {Bennert}, {Lintott},
  {Chojnowski}, {Moiseev}, {Smirnova}, {Schawinski}, {Urry}, {Evans},
  {Pancoast}, {Scott}, {Showley}, \& {Flatland}}]{KE15.1}
{Keel}, W.~C., {Maksym}, W.~P., {Bennert}, V.~N., {Lintott}, C.~J.,
  {Chojnowski}, S.~D., {Moiseev}, A., {Smirnova}, A., {Schawinski}, K., {Urry},
  C.~M., {Evans}, D.~A., {Pancoast}, A., {Scott}, B., {Showley}, C., \&
  {Flatland}, K. 2015, \aj, 149, 155

\bibitem[{{Kennicutt} \& {Evans}(2012)}]{KE12.1}
{Kennicutt}, R.~C., \& {Evans}, N.~J. 2012, \araa, 50, 531

\bibitem[{{Kewley} {et~al.}(2001){Kewley}, {Dopita}, {Sutherland}, {Heisler},
  \& {Trevena}}]{KE01.2}
{Kewley}, L.~J., {Dopita}, M.~A., {Sutherland}, R.~S., {Heisler}, C.~A., \&
  {Trevena}, J. 2001, \apj, 556, 121

\bibitem[{{Kewley} {et~al.}(2006){Kewley}, {Groves}, {Kauffmann}, \&
  {Heckman}}]{KE06.1}
{Kewley}, L.~J., {Groves}, B., {Kauffmann}, G., \& {Heckman}, T. 2006, \mnras,
  372, 961

\bibitem[{{Koyama} {et~al.}(1996){Koyama}, {Maeda}, {Sonobe}, {Takeshima},
  {Tanaka}, \& {Yamauchi}}]{KO96.2}
{Koyama}, K., {Maeda}, Y., {Sonobe}, T., {Takeshima}, T., {Tanaka}, Y., \&
  {Yamauchi}, S. 1996, \pasj, 48, 249

\bibitem[{{Krajnovi{\'c}} {et~al.}(2006){Krajnovi{\'c}}, {Cappellari}, {de
  Zeeuw}, \& {Copin}}]{KR06.2}
{Krajnovi{\'c}}, D., {Cappellari}, M., {de Zeeuw}, P.~T., \& {Copin}, Y. 2006,
  \mnras, 366, 787

\bibitem[{Lagarias {et~al.}(1998)Lagarias, Reeds, Wright, \& Wright}]{LA98.1}
Lagarias, J.~C., Reeds, J.~A., Wright, M.~H., \& Wright, P.~E. 1998, SIAM
  Journal on optimization, 9, 112

\bibitem[{{LaMassa} {et~al.}(2015){LaMassa}, {Cales}, {Moran}, {Myers},
  {Richards}, {Eracleous}, {Heckman}, {Gallo}, \& {Urry}}]{LA15.2}
{LaMassa}, S.~M., {Cales}, S., {Moran}, E.~C., {Myers}, A.~D., {Richards},
  G.~T., {Eracleous}, M., {Heckman}, T.~M., {Gallo}, L., \& {Urry}, C.~M. 2015,
  \apj, 800, 144

\bibitem[{{Lintott} {et~al.}(2009){Lintott}, {Schawinski}, {Keel}, {van Arkel},
  {Bennert}, {Edmondson}, {Thomas}, {Smith}, {Herbert}, {Jarvis}, {Virani},
  {Andreescu}, {Bamford}, {Land}, {Murray}, {Nichol}, {Raddick}, {Slosar},
  {Szalay}, \& {Vandenberg}}]{LI09.2}
{Lintott}, C.~J., {Schawinski}, K., {Keel}, W., {van Arkel}, H., {Bennert}, N.,
  {Edmondson}, E., {Thomas}, D., {Smith}, D.~J.~B., {Herbert}, P.~D., {Jarvis},
  M.~J., {Virani}, S., {Andreescu}, D., {Bamford}, S.~P., {Land}, K., {Murray},
  P., {Nichol}, R.~C., {Raddick}, M.~J., {Slosar}, A., {Szalay}, A., \&
  {Vandenberg}, J. 2009, \mnras, 399, 129

\bibitem[{{Liu} {et~al.}(2010){Liu}, {Shen}, {Strauss}, \& {Greene}}]{LI10.1}
{Liu}, X., {Shen}, Y., {Strauss}, M.~A., \& {Greene}, J.~E. 2010, \apj, 708,
  427

\bibitem[{{Liu} {et~al.}(2011){Liu}, {Shen}, {Strauss}, \& {Hao}}]{LI11.3}
{Liu}, X., {Shen}, Y., {Strauss}, M.~A., \& {Hao}, L. 2011, \apj, 737, 101

\bibitem[{{Malkan} {et~al.}(1998){Malkan}, {Gorjian}, \& {Tam}}]{MA98.1}
{Malkan}, M.~A., {Gorjian}, V., \& {Tam}, R. 1998, \apjs, 117, 25

\bibitem[{{Martini} {et~al.}(2003){Martini}, {Regan}, {Mulchaey}, \&
  {Pogge}}]{MA03.6}
{Martini}, P., {Regan}, M.~W., {Mulchaey}, J.~S., \& {Pogge}, R.~W. 2003,
  \apjs, 146, 353

\bibitem[{{McGurk} {et~al.}(2015){McGurk}, {Max}, {Medling}, {Shields}, \&
  {Comerford}}]{MC15.1}
{McGurk}, R.~C., {Max}, C.~E., {Medling}, A.~M., {Shields}, G.~A., \&
  {Comerford}, J.~M. 2015, \apj, 811, 14

\bibitem[{{Moy} \& {Rocca-Volmerange}(2002)}]{MO02.1}
{Moy}, E., \& {Rocca-Volmerange}, B. 2002, \aap, 383, 46

\bibitem[{{Mulchaey} {et~al.}(1996){Mulchaey}, {Wilson}, \&
  {Tsvetanov}}]{MU96.1}
{Mulchaey}, J.~S., {Wilson}, A.~S., \& {Tsvetanov}, Z. 1996, \apjs, 102, 309

\bibitem[{{M{\"u}ller-S{\'a}nchez} {et~al.}(2016){M{\"u}ller-S{\'a}nchez},
  {Comerford}, {Stern}, \& {Harrison}}]{MU16.1}
{M{\"u}ller-S{\'a}nchez}, F., {Comerford}, J., {Stern}, D., \& {Harrison},
  F.~A. 2016, \apj, 830, 50

\bibitem[{{M{\"u}ller-S{\'a}nchez} {et~al.}(2015){M{\"u}ller-S{\'a}nchez},
  {Comerford}, {Nevin}, {Barrows}, {Cooper}, \& {Greene}}]{MU15.1}
{M{\"u}ller-S{\'a}nchez}, F., {Comerford}, J.~M., {Nevin}, R., {Barrows},
  R.~S., {Cooper}, M.~C., \& {Greene}, J.~E. 2015, \apj, 813, 103

\bibitem[{{M{\"u}ller S{\'a}nchez} {et~al.}(2009){M{\"u}ller S{\'a}nchez},
  {Davies}, {Genzel}, {Tacconi}, {Eisenhauer}, {Hicks}, {Friedrich}, \&
  {Sternberg}}]{MU09.1}
{M{\"u}ller S{\'a}nchez}, F., {Davies}, R.~I., {Genzel}, R., {Tacconi}, L.~J.,
  {Eisenhauer}, F., {Hicks}, E.~K.~S., {Friedrich}, S., \& {Sternberg}, A.
  2009, \apj, 691, 749

\bibitem[{{Nandra} \& {Pounds}(1994)}]{NA94.2}
{Nandra}, K., \& {Pounds}, K.~A. 1994, \mnras, 268, 405

\bibitem[{{Nevin} {et~al.}(2016){Nevin}, {Comerford}, {M{\"u}ller-S{\'a}nchez},
  {Barrows}, \& {Cooper}}]{NE16.1}
{Nevin}, R., {Comerford}, J., {M{\"u}ller-S{\'a}nchez}, F., {Barrows}, R., \&
  {Cooper}, M. 2016, \apj, 832, 67

\bibitem[{{Nevin} {et~al.}(2017){Nevin}, {Comerford}, {M{\"u}ller S{\'a}nchez},
  {Barrows}, \& {Cooper}}]{NE17.1}
{Nevin}, R., {Comerford}, J.~M., {M{\"u}ller S{\'a}nchez}, F., {Barrows},
  R.~S., \& {Cooper}, M.~C. 2017, MNRAS submitted

\bibitem[{{Norman} \& {Ikeuchi}(1989)}]{NO89.1}
{Norman}, C.~A., \& {Ikeuchi}, S. 1989, \apj, 345, 372

\bibitem[{{Oh} {et~al.}(2011){Oh}, {Sarzi}, {Schawinski}, \& {Yi}}]{OH11.1}
{Oh}, K., {Sarzi}, M., {Schawinski}, K., \& {Yi}, S.~K. 2011, \apjs, 195, 13

\bibitem[{{Oh} {et~al.}(2015){Oh}, {Yi}, {Schawinski}, {Koss}, {Trakhtenbrot},
  \& {Soto}}]{OH15.1}
{Oh}, K., {Yi}, S.~K., {Schawinski}, K., {Koss}, M., {Trakhtenbrot}, B., \&
  {Soto}, K. 2015, \apjs, 219, 1

\bibitem[{{Osterbrock} \& {Ferland}(2006)}]{OS06.1}
{Osterbrock}, D.~E., \& {Ferland}, G.~J. 2006

\bibitem[{{Park} {et~al.}(2006){Park}, {Kashyap}, {Siemiginowska}, {van Dyk},
  {Zezas}, {Heinke}, \& {Wargelin}}]{PA06.1}
{Park}, T., {Kashyap}, V.~L., {Siemiginowska}, A., {van Dyk}, D.~A., {Zezas},
  A., {Heinke}, C., \& {Wargelin}, B.~J. 2006, \apj, 652, 610

\bibitem[{{Peng} {et~al.}(2010){Peng}, {Ho}, {Impey}, \& {Rix}}]{PE10.1}
{Peng}, C.~Y., {Ho}, L.~C., {Impey}, C.~D., \& {Rix}, H.-W. 2010, \aj, 139,
  2097

\bibitem[{{Piconcelli} {et~al.}(2005){Piconcelli}, {Jimenez-Bail{\'o}n},
  {Guainazzi}, {Schartel}, {Rodr{\'{\i}}guez-Pascual}, \&
  {Santos-Lle{\'o}}}]{PI05.1}
{Piconcelli}, E., {Jimenez-Bail{\'o}n}, E., {Guainazzi}, M., {Schartel}, N.,
  {Rodr{\'{\i}}guez-Pascual}, P.~M., \& {Santos-Lle{\'o}}, M. 2005, \aap, 432,
  15

\bibitem[{{Reeves} \& {Turner}(2000)}]{RE00.1}
{Reeves}, J.~N., \& {Turner}, M.~J.~L. 2000, \mnras, 316, 234

\bibitem[{{Richardson} {et~al.}(2014){Richardson}, {Allen}, {Baldwin},
  {Hewett}, \& {Ferland}}]{RI14.2}
{Richardson}, C.~T., {Allen}, J.~T., {Baldwin}, J.~A., {Hewett}, P.~C., \&
  {Ferland}, G.~J. 2014, \mnras, 437, 2376

\bibitem[{{Runnoe} {et~al.}(2016){Runnoe}, {Cales}, {Ruan}, {Eracleous},
  {Anderson}, {Shen}, {Green}, {Morganson}, {LaMassa}, {Greene}, {Dwelly},
  {Schneider}, {Merloni}, {Georgakakis}, \& {Roman-Lopes}}]{RU16.1}
{Runnoe}, J.~C., {Cales}, S., {Ruan}, J.~J., {Eracleous}, M., {Anderson},
  S.~F., {Shen}, Y., {Green}, P.~J., {Morganson}, E., {LaMassa}, S., {Greene},
  J.~E., {Dwelly}, T., {Schneider}, D.~P., {Merloni}, A., {Georgakakis}, A., \&
  {Roman-Lopes}, A. 2016, \mnras, 455, 1691

\bibitem[{{Schawinski} {et~al.}(2015){Schawinski}, {Koss}, {Berney}, \&
  {Sartori}}]{SC15.1}
{Schawinski}, K., {Koss}, M., {Berney}, S., \& {Sartori}, L.~F. 2015, \mnras,
  451, 2517

\bibitem[{{Schirmer} {et~al.}(2016){Schirmer}, {Malhotra}, {Levenson}, {Fu},
  {Davies}, {Keel}, {Torrey}, {Bennert}, {Pancoast}, \& {Turner}}]{SC16.2}
{Schirmer}, M., {Malhotra}, S., {Levenson}, N.~A., {Fu}, H., {Davies}, R.~L.,
  {Keel}, W.~C., {Torrey}, P., {Bennert}, V.~N., {Pancoast}, A., \& {Turner},
  J.~E.~H. 2016, \mnras, 463, 1554

\bibitem[{{Schmitt} {et~al.}(2003){Schmitt}, {Donley}, {Antonucci},
  {Hutchings}, \& {Kinney}}]{SC03.3}
{Schmitt}, H.~R., {Donley}, J.~L., {Antonucci}, R.~R.~J., {Hutchings}, J.~B.,
  \& {Kinney}, A.~L. 2003, \apjs, 148, 327

\bibitem[{{Schnorr-M{\"u}ller} {et~al.}(2016){Schnorr-M{\"u}ller},
  {Storchi-Bergmann}, {Robinson}, {Lena}, \& {Nagar}}]{SC16.1}
{Schnorr-M{\"u}ller}, A., {Storchi-Bergmann}, T., {Robinson}, A., {Lena}, D.,
  \& {Nagar}, N.~M. 2016, \mnras, 457, 972

\bibitem[{{Shopbell} \& {Bland-Hawthorn}(1998)}]{SH98.1}
{Shopbell}, P.~L., \& {Bland-Hawthorn}, J. 1998, \apj, 493, 129

\bibitem[{{Stern} {et~al.}(2012){Stern}, {Assef}, {Benford}, {Blain}, {Cutri},
  {Dey}, {Eisenhardt}, {Griffith}, {Jarrett}, {Lake}, {Masci}, {Petty},
  {Stanford}, {Tsai}, {Wright}, {Yan}, {Harrison}, \& {Madsen}}]{ST12.2}
{Stern}, D., {Assef}, R.~J., {Benford}, D.~J., {Blain}, A., {Cutri}, R., {Dey},
  A., {Eisenhardt}, P., {Griffith}, R.~L., {Jarrett}, T.~H., {Lake}, S.,
  {Masci}, F., {Petty}, S., {Stanford}, S.~A., {Tsai}, C.-W., {Wright}, E.~L.,
  {Yan}, L., {Harrison}, F., \& {Madsen}, K. 2012, \apj, 753, 30

\bibitem[{{Su} {et~al.}(2010){Su}, {Slatyer}, \& {Finkbeiner}}]{SU10.1}
{Su}, M., {Slatyer}, T.~R., \& {Finkbeiner}, D.~P. 2010, \apj, 724, 1044

\bibitem[{{Sun} {et~al.}(2016){Sun}, {Greene}, \& {Zakamska}}]{SU16.1}
{Sun}, A.-L., {Greene}, J.~E., \& {Zakamska}, N.~L. 2016, ArXiv e-prints

\bibitem[{{Sun} {et~al.}(2014){Sun}, {Greene}, {Zakamska}, \&
  {Nesvadba}}]{SU14.1}
{Sun}, A.-L., {Greene}, J.~E., {Zakamska}, N.~L., \& {Nesvadba}, N.~P.~H. 2014,
  \apj, 790, 160

\bibitem[{{Sutherland} {et~al.}(1993){Sutherland}, {Bicknell}, \&
  {Dopita}}]{SU93.1}
{Sutherland}, R.~S., {Bicknell}, G.~V., \& {Dopita}, M.~A. 1993, \apj, 414, 510

\bibitem[{{Tremonti} {et~al.}(2007){Tremonti}, {Moustakas}, \&
  {Diamond-Stanic}}]{TR07.1}
{Tremonti}, C.~A., {Moustakas}, J., \& {Diamond-Stanic}, A.~M. 2007, \apjl,
  663, L77

\bibitem[{{Veilleux} {et~al.}(2005){Veilleux}, {Cecil}, \&
  {Bland-Hawthorn}}]{VE05.1}
{Veilleux}, S., {Cecil}, G., \& {Bland-Hawthorn}, J. 2005, \araa, 43, 769

\bibitem[{{Veilleux} \& {Osterbrock}(1987)}]{VE87.1}
{Veilleux}, S., \& {Osterbrock}, D.~E. 1987, \apjs, 63, 295

\bibitem[{{Wilson} \& {Tsvetanov}(1994)}]{WI94.1}
{Wilson}, A.~S., \& {Tsvetanov}, Z.~I. 1994, \aj, 107, 1227

\bibitem[{{Zhang} {et~al.}(2015){Zhang}, {Hailey}, {Mori}, {Clavel}, {Terrier},
  {Ponti}, {Goldwurm}, {Bauer}, {Boggs}, {Christensen}, {Craig}, {Harrison},
  {Hong}, {Nynka}, {Soldi}, {Stern}, {Tomsick}, \& {Zhang}}]{ZH15.1}
{Zhang}, S., {Hailey}, C.~J., {Mori}, K., {Clavel}, M., {Terrier}, R., {Ponti},
  G., {Goldwurm}, A., {Bauer}, F.~E., {Boggs}, S.~E., {Christensen}, F.~E.,
  {Craig}, W.~W., {Harrison}, F.~A., {Hong}, J., {Nynka}, M., {Soldi}, S.,
  {Stern}, D., {Tomsick}, J.~A., \& {Zhang}, W.~W. 2015, \apj, 815, 132

\end{thebibliography}

\end{document}